%% file: main.tex
\newif\iftr\trfalse
\trtrue

\documentclass[conference]{IEEEtran}

\makeatletter
\def\endthebibliography{%
  \def\@noitemerr{\@latex@warning{Empty `thebibliography' environment}}%
  \endlist
}
\makeatother

\IEEEpeerreviewmaketitle

\usepackage{filecontents}
\usepackage[noadjust]{cite}
\usepackage{amsmath}
\usepackage{amssymb}
\usepackage{ebproof}
\usepackage{todonotes}
\usepackage{amsfonts}
\usepackage{algorithmic}
\usepackage{textcomp}
\usepackage{mathpartir}
\usepackage{color}
\usepackage{xcolor}
\DeclareMathAlphabet{\mathpzc}{OT1}{pzc}{m}{it}

\usepackage{subcaption} 
\usepackage{epsfig}
\usepackage{listings}
\usepackage[override]{cmtt}
\usepackage{graphicx}
\usepackage{balance}
\usepackage{xspace}
\usepackage{booktabs}
\usepackage{url}
\usepackage{xparse}
\usepackage{enumitem}
\usepackage{tikz}
\usetikzlibrary{chains,fit,shapes,positioning,calc,arrows.meta,shapes.arrows}

\usepackage[numbers]{natbib}
\lstdefinestyle{customc}{
  belowcaptionskip=1\baselineskip,
  breaklines=true,
  breakatwhitespace, 
  numbers=left,
  xleftmargin=\parindent,
  language=C,
  columns=flexible,      
  showstringspaces=false,
  basicstyle=\small\sffamily,
  otherkeywords={uint,uintptr_t,let,assert},
  literate={{<-}{{$\leftarrow\,$}}2
            {->}{{$\rightarrow\,$}}2
            {<=}{{$\leq\,$}}2},
  numberstyle=\tiny\rmfamily,
  keywordstyle=\bfseries\color{green!40!black},
  commentstyle=\itshape\color{purple!40!black},
  identifierstyle=\color{blue!80!black},
}
\newenvironment{DIFnomarkup}{}{}
\newcommand{\myparagraph}[1]{\textbf{#1}.\xspace}

\newcommand{\code}[1]{\lstinline|#1|}

\newif\ifsubmit\submitfalse

\ifsubmit
\newcommand{\mwh}[1]{}
\newcommand{\ashe}[1]{}
\newcommand{\dtarditi}[1]{}
\newcommand{\dvh}[1]{}
\newcommand{\leo}[1]{}
\newcommand{\liyi}[1]{}
\newcommand{\yiyun}[1]{}
\newcommand{\review}[1]{}

\else
\newcommand{\mwh}[1]{\textcolor{red}{Mike: #1}}
\newcommand{\dtarditi}[1]{\textcolor{purple}{David: #1}}
\newcommand{\leo}[1]{\textcolor{green!80!blue}{Leo: #1}}
\newcommand{\liyi}[1]{\textbf{\textcolor{orange}{Liyi: #1}}}
\newcommand{\yiyun}[1]{\textcolor{cyan}{Yiyun: #1}}
\newcommand{\dvh}[1]{\textcolor{magenta}{DVH: #1}}
\newcommand{\review}[1]{\textbf{\textcolor{blue}{Review: #1}}}

\fi

\newcommand{\lang}{\textsc{CoreChkC}\xspace}
\newcommand{\elang}{\textsc{CoreC}\xspace}

\newtheorem{defi}{Definition}
\newtheorem{thm}{Theorem}

\newcommand{\CoreChkC}{\lang}
\newcommand{\checkedc}{\text{Checked C}\xspace}

\newcommand{\tnat}{\texttt{nat}}

\newcommand{\kw}[1]{\ensuremath{\mathtt{#1}}}

\newcommand{\estrlen}[1]{\ensuremath{\kw{strlen}({#1})}}
\newcommand{\estrlentext}{\ensuremath{\kw{strlen}}}

\newcommand{\tarray}[3]{\tarrayb{({#1},{#2})}{#3}}
\newcommand{\tarrayb}[2]{\ensuremath{[{#1}~{#2}]}}
\newcommand{\tntarray}[3]{\tntarrayb{({#1},{#2})}{#3}}
\newcommand{\tntarrayb}[2]{\tarrayb{#1}{#2}_{nt}}
\newcommand{\tallarrayb}[2]{\ensuremath{[{#1}~{#2}]_{\kappa}}}
\newcommand{\tallarraybc}[2]{\ensuremath{[{#1}~{#2}]_{\textcolor{cyan}{\kappa}}}}

\newcommand{\tallarray}[3]{\tallarrayb{({#1},{#2})}{#3}}

\newcommand{\tptr}[2]{\ensuremath{\mathtt{ptr}^{#2}~{#1}}}

\newcommand{\tarrayptr}[4]{{\tptr{\tarray{#1}{#2}{#3}}{#4}}}
\newcommand{\tntarrayptr}[4]{{\tptr{\tntarray{#1}{#2}{#3}}{#4}}}
\newcommand{\tallarrayptr}[4]{{\tptr{\tallarray{#1}{#2}{#3}}{#4}}}

\newcommand{\tstruct}[1]{\ensuremath{\kw{struct}~{#1}}}

\newcommand{\evalue}[2]{\ensuremath{{#1}\!:\!{#2}}}

\newcommand{\emalloc}[1]{\ensuremath{\kw{malloc}({#1})}}
\newcommand{\emalloctext}{\ensuremath{\kw{malloc}}}
\newcommand{\ecall}[2]{\ensuremath{{#1}({#2})}}
\newcommand{\ret}[3]{\ensuremath{\kw{ret}({#1},{#2},{#3})}}

\newcommand{\ecast}[2]{\ensuremath{\kw{(}{#1}\kw{)}{#2}}}
\newcommand{\edyncast}[2]{\ensuremath{\langle{#1}\rangle{#2}}}

\newcommand{\elet}[3]{\ensuremath{\kw{let}~#1\, \texttt{=}\, #2~\kw{in}\;{#3}}}
\newcommand{\elettext}{\ensuremath{\kw{let}}}
\newcommand{\ebinop}[2]{\ensuremath{#1 \plus #2}}
\newcommand{\eassign}[2]{\ensuremath{\texttt{*}{#1}\,\texttt{=}\, {#2}}}
\newcommand{\eassignstack}[2]{\ensuremath{{#1}\,\texttt{=}\, {#2}}}

\newcommand{\estar}[1]{\ensuremath{\texttt{*}{#1}}}
\newcommand{\eamper}[2]{\ensuremath{\kw{\&}{#1}\kw{\rightarrow}{#2}}}
\newcommand{\eunchecked}[1]{\ensuremath{\kw{unchecked}\;{#1}}}
\newcommand{\euncheckedtext}{\ensuremath{\kw{unchecked}}}
\newcommand{\eif}[3]{\ensuremath{\kw{if\;}(#1)\;{#2}\;\kw{else}\;{#3}}}

\newcommand{\eiftext}{\ensuremath{\kw{if}}}
\newcommand{\ethentext}{\ensuremath{\kw{then}}}

\newcommand{\ebounds}{\ensuremath{\kw{bounds}}}
\newcommand{\enull}{\ensuremath{\kw{null}}}
\newcommand{\hole}{\ensuremath{\Box}}
\newcommand{\defscope}{\sigma}

\newcommand{\tint}{\ensuremath{\mathtt{int}}}

\newcommand{\efalse}{\ensuremath{\mathtt{false}}}

\newcommand{\heap}{\ensuremath{\mathpzc{H}}}
\newcommand{\eret}[3]{\ensuremath{\kw{ret}({#1},{#2},{#3})}}
\newcommand{\erettext}{\ensuremath{\kw{ret}}}

\newcommand{\plus}{\mathbin{\texttt{+}}}

\newcommand{\fv}{\mathit{FV}}

\newcommand{\size}{\mathit{size}}

\newcommand{\cmode}{\texttt{c}}
\newcommand{\umode}{\texttt{u}}
\newcommand{\bvar}{\ensuremath{\beta}}
\newcommand{\mode}{\textit{mode}}

\newcommand{\cextend}[4]{\ensuremath{#4 = ~ \vdash_{extend} #1, #2, #3}}
\newcommand{\fresh}{\ensuremath{\kw{fresh}}}
\newcommand{\echecknull}[3]{\ensuremath{#3 = ~\vdash_{null}#1, #2}}
\newcommand{\echeckboundsdyn}[5]{\ensuremath{#5 = ~ \vdash_{boundsD}#1, #2, #3, #4}}
\newcommand{\echeckbounds}[5]{\ensuremath{#5 = ~ \vdash_{boundsR}#1, #2, #3, #4}}
\newcommand{\echeckboundsw}[5]{\ensuremath{#5 = ~ \vdash_{boundsW}#1, #2, #3, #4}}
\newcommand{\esizeof}[1]{\ensuremath{\kw{sizeof}(#1)}}
\newcommand{\ewidenstrlen}[5]{\ensuremath{#5 = ~ \vdash_{widenstr}#1, #2, #3, #4}}
\newcommand{\ewidenderef}[4]{\ensuremath{#4 = ~ \vdash_{widenderef}#1, #2, #3}}

\definecolor{programs}{gray}{0.1}

\makeatletter
\lst@CCPutMacro
    \lst@ProcessOther {"22}{\lst@ifupquote \textquotedbl
                                     \else \char34\relax \fi}
    \@empty\z@\@empty
\makeatother

\def\BibTeX{{\rm B\kern-.05em{\sc i\kern-.025em b}\kern-.08em
    T\kern-.1667em\lower.7ex\hbox{E}\kern-.125emX}}
\begin{document}

\iftr
\title{A Formal Model of Checked C {\large (Extended Version)}}
\else
\title{A Formal Model of Checked C}
\fi


\author{Liyi Li, Yiyun Liu$^\dagger$, Deena Postol, Leonidas
  Lampropoulos, David Van Horn, and Michael Hicks\\
  University of Maryland $\quad\quad ~^\dagger$University of Pennsylvania}
  


\maketitle

\begin{abstract}

  We present a formal model of Checked C, a dialect of C
  that aims to enforce spatial memory safety. Our model pays particular
  attention to the semantics of dynamically sized, potentially null-terminated
  arrays.
  We formalize this model in Coq, and prove that any spatial memory
  safety errors can be \emph{blamed} on portions of the program
  labeled \emph{unchecked}; this is a Checked C feature that supports
  incremental porting and backward compatibility.
While our model's operational semantics uses annotated
  (``fat'') pointers to enforce spatial
  safety, we show that such annotations can be safely erased.
    Using PLT Redex we formalize an executable
    version of our model and a compilation procedure to an untyped C-like
    language, as well as use randomized testing to
    validate that generated code faithfully simulates the original.
  Finally, we develop a custom random generator 
  for well-typed and almost-well-typed terms in our Redex model, and
  use it to search for
  inconsistencies between our model and the Clang Checked C
  implementation. We find these steps to be a useful way to co-develop a language
  (Checked C is still in development) and a core model of it.
\end{abstract}

\iftr
\bigskip
\begin{small}
 This is an extended version of a paper that
  appears at the 2022 Computer Security Foundations Symposium.
\end{small}
\fi
  
\input{introduction}

\input{background}

\input{formal.tex}

\input{compilation.tex}

\input{evaluation}
\input{comparison}

\input{conclusion}

\bigskip

\paragraph*{Acknowledgments}
We thank the anonymous reviewers for their helpful, constructive
comments. This work was supported in part by a gift from Microsoft. 




\begin{small}
\balance
\bibliography{IEEEabrv,paper,sources}
\end{small}

\iftr
\newpage
\appendix
\input{appendix}

\fi

\end{document}

%% file: introduction.tex




\section{Introduction}\label{sec:intros}



The C programming language remains extremely popular despite the
emergence of new, modern languages. Unfortunately, C programs lack
spatial memory safety, which makes them susceptible to a host
of devastating vulnerabilities, including buffer overflows and
out-of-bounds reads/writes. Despite their long history, buffer
overflows and other spatial safety violations are among the most
prevalent and dangerous vulnerabilities on the Internet today \cite{Zeng:2013:SRF:2534766.2534798}.


Several industrial and research efforts---including CCured~\cite{Necula2005},
Softbound~\cite{softbound}, and ASAN~\cite{Serebryany2012}---have
explored means to compile C programs 
to automatically enforce spatial safety. These
approaches all impose performance overheads deemed too high for
deployment use. Recently, \citet{Elliott2018} introduced \checkedc, an
open-source extension to C with new types and
annotations whose use can ensure a program’s spatial safety.
Importantly, \checkedc supports development that is 
incremental and compositional. Code regions (e.g.,
functions or whole files) designated as \emph{checked} enforce
spatial safety in a manner preserved by composition with
other checked regions. But not all regions must be checked: Checked
C's annotated \emph{checked pointers} are binary-compatible with legacy pointers, and
may coexist in the same code, which permits a deliberate (and
semi-automated) refactoring process. Parts of the FreeBSD kernel have
been successfully ported to \checkedc~\cite{duanrefactoring}, and overall, performance
overhead seems low enough for practical deployment.

While \checkedc promises to enforce spatial safety, we might wonder
whether its design and implementation deliver on this promise, or even
what ``spatial safety'' means when a program contains both checked and
unchecked code. In prior work, \citet{ruef18checkedc-incr} developed a core
formalization of \checkedc and with it proved a \emph{soundness} theorem for
checked code: any stuck (i.e., ill-defined)
state reached by a well-typed program amounts to a spatial safety
violation; such a state can always be attributed to, i.e., 
\emph{blamed on}, the execution of code that is not in a checked
region. While their work is a good start, 
it fails to model important aspects of \checkedc's functionality,
particularly those involving pointers to arrays. In this paper, we
cover this gap, making three main contributions.

\myparagraph{Dynamically bounded and null-terminated arrays}
Our first contribution is a core formalism called \lang, which extends
\citet{ruef18checkedc-incr} with several new features, most notably 
\emph{dynamically bounded arrays} (Section~\ref{sec:formal}). 
Dynamically bounded arrays are those whose size is
known only at run time, as designated by in-scope variables using
dependent types. A pointer's accessible memory is bounded both above
and below, to admit arbitrary pointer arithmetic.

\lang also models
\emph{null-terminated} arrays, a kind of dynamically bounded array
whose upper bound
defines the array's \emph{minimum} length---additional space is available 
up to a null terminator. For example, the \checkedc type
\lstinline{nt_array_ptr<char> p:count(n)} says that
\lstinline{p} has length \emph{at least} \lstinline{n} (excluding the
null terminator), but further capacity is present if %
\lstinline|p[n]| is not null. \checkedc (and \lang) supports flow-sensitive
\emph{bounds widening}: statements of the form \lstinline|if (*p) | $s$,
where \lstinline{p}'s type is \lstinline{nt_array_ptr<T> count(0)},
typecheck statement $s$ under the assumption that \lstinline{p} has
type \lstinline{nt_array_ptr<T> count(1)}, 
 i.e., one more than it was,
since the character at the current-known length is
non-null. Similarly, the call \lstinline|n = strlen(p)| will widen
\lstinline|p|'s bounds to \lstinline|n|. Subtyping permits treating
null-terminated arrays as normal arrays of the same size (which
does not include, and thereby protects, the null
terminator).\footnote{
    See Sec.~\ref{sec:related} for a careful comparison of
    \citet{ruef18checkedc-incr} and \lang.
}

We prove, in Coq, a blame theorem for \lang.  As far as we are aware,
ours is the first formalized type system and proof of soundness for
pointers to null-terminated arrays with expandable bounds.


\myparagraph{Sound compilation of checked pointers} Our second
contribution is a formalization of bounds-check insertion for array
accesses (Section~\ref{sec:compilation}). Our operational semantics
annotates each pointer with metadata that describes its bounds, and
the assignment and dereference rules have premises to confirm the
access is in bounds. An obvious compilation scheme (taken by
Cyclone~\cite{Jim2002,GrossmanMJHWC02}, CCured~\cite{Necula2005}, and
earlier works) would be to translate annotated pointers to multi-word
objects: one word for the pointer, and 1-2 words to describe its lower
and upper bounds. Inserted checks reference these bounds. While
convenient, such ``fat'' pointers are expensive, and break backward
binary compatibility with legacy pointers.

  To show that pointer annotations can be safely
  erased, and thus fat pointers are not needed, we formalize a
  translation of \lang to \elang, which is an 
  untyped version of \lang that drops metadata annotations, and
  lacks bounds/null checks in the semantics rules. Instead,
  the compilation process inserts null/bounds checks explicitly, leveraging
  compile-time type information. While we do not definitively prove
  it, we provide strong evidence that compilation is correct. We use PLT
  Redex~\cite{pltredex} to mechanize (a generalization of) \lang, 
  \elang, and compilation between the two, and we use randomized testing 
to validate that the compiled program \emph{simulates} the
original. In addition to demonstrating the technical point that metadata
annotations in the \lang formalism do not necessitate fat pointers,
compilation also sheds light on the actual Checked C compilation
process. 

As far as we are aware, \lang is the first formalism to cleanly
separate bounds-checking compilation from the core semantics; prior
work~\cite{Feng2006,Condit2007} merged the two, conflating
\emph{meaning} with \emph{mechanism}. In carrying out the
formalization, we discovered that our compilation approach for
null-terminated array pointers is more expressive than that proposed
in the Checked C specification~\cite{checkedc}
(Section~\ref{sec:disc}); we would not have discovered this
improvement had we not separated checks from semantics.

\myparagraph{Model-based randomized testing} Our third and final
contribution is a strategy and implementation of model-based
randomized testing (Section~\ref{sec:evaluation}).  To check the
correctness of our formal model, we compare the behavior between the
existing Clang \checkedc implementation and our own model. This is
done by a conversion tool that converts expressions from \lang into
actual \checkedc code that can be compiled by the Clang \checkedc
compiler. We build a random generator of programs largely based on the
typing rules of \lang and make sure that, both statically and
dynamically, \lang and Clang Checked C are consistent after
conversion.  This helped rapidly prototype the model and uncovered
several issues in the Checked C compiler.

\begin{figure}
\includegraphics[width=0.5\textwidth]{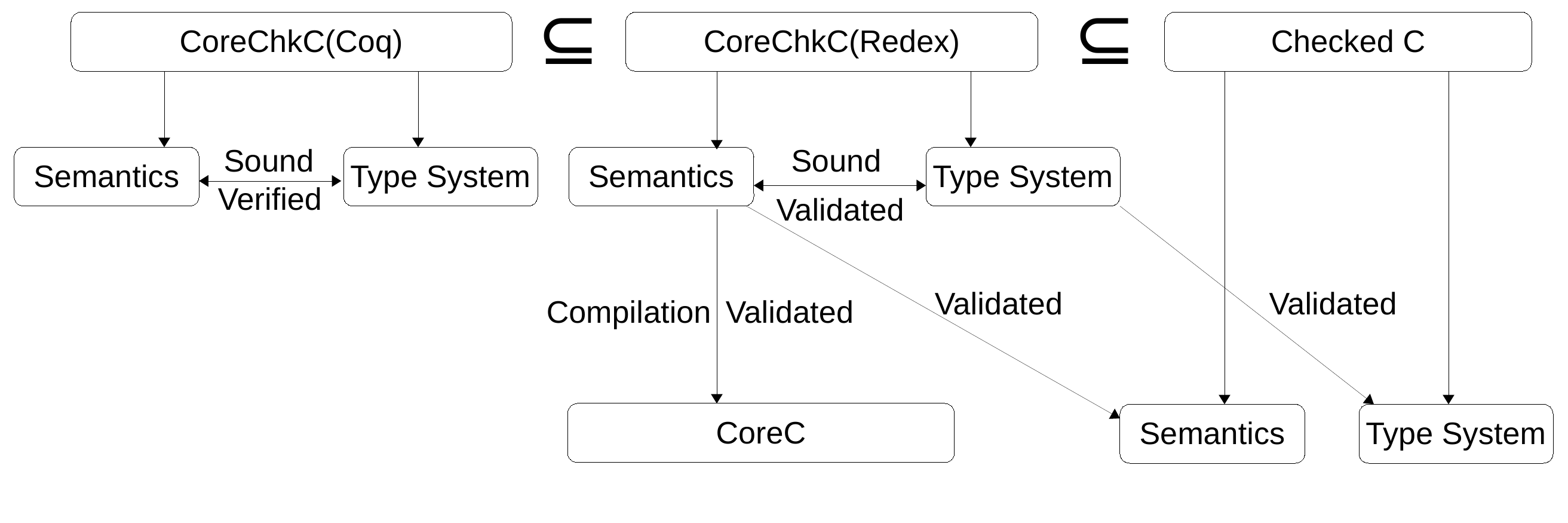}
\caption{
    \lang models' relationship to \checkedc
}
  \label{fig:model-relation}
\end{figure}

  \myparagraph{Summary Visualization} The
  relationship among our contributions is visualized in
  Fig.~\ref{fig:model-relation}. With the Coq model of \lang we prove
  soundness (and with it, \emph{blame}) of the \checkedc type system
  and semantics. With the Redex model, we use randomized testing to
  validate both type soundness and compilation correctness, where the
  latter shows how compilation need not output fat pointers despite
  the use of pointer annotations in the \lang model. The Redex \lang
  model is also the basis of randomized testing of the correctness of
  the \checkedc compiler implementation, both its type checker and the
  semantics of its emitted code, at least for the subset of the
  language in the Redex model. The Redex model's syntax
  is slightly richer than the Coq version: conditional guards and
  function arguments may be arbitrary expressions, where the Coq
  version limits them to constants and variables, making handling of
  dependent types a bit simpler. We find a useful synergy between the
  Coq and Redex models for carrying out a language development. The
  richer, executable Redex model is useful for quickly modeling and
  testing new features, both formally and against a real
  implementation. Once solidified, new features can be added to the
  Coq model (perhaps somewhat simplified) for final proofs of
  correctness.

We begin with a review of \checkedc (Section~\ref{sec:overview}),
present our main contributions
(Sections~\ref{sec:formal}--\ref{sec:evaluation}), and conclude with a
discussion of 
related and future work (Sections~\ref{sec:related},
\ref{sec:conclude}). All code and proof artifacts (both for Coq and
Redex) can be found at \url{https://github.com/plum-umd/checkedc}.

%% file: background.tex
\section{\checkedc Overview}\label{sec:overview}

This section describes Checked C, which extends C with new pointer
types and annotations that ensure spatial safety. Development of
Checked C was initiated by Microsoft Research in 2015 but starting in
late 2021 was forked and is now actively managed by the Secure Software
Development Project (SSDP). Details can be found in a prior
overview~\cite{Elliott2018} or the full specification~\cite{checkedc}.
Checked C is implemented as an extension of Clang/LLVM and the SSDP
fork is freely available at \url{https://github.com/secure-sw-dev}. 

\begin{figure}[t]
{\small
  \begin{lstlisting}[xleftmargin=4 mm]
nt_array_ptr<const char>
parse_utf16_hex(nt_array_ptr<const char> s,
                  ptr<uint> result) {
 int x1, x2, x3, x4;
 if (s[0] != 0) { x1 = hex_char_to_int(s[0]);
 if (s[1] != 0) { x2 = hex_char_to_int(s[1]);
 if (s[2] != 0) { x3 = hex_char_to_int(s[2]);
 if (s[3] != 0) { x4 = hex_char_to_int(s[3]);
 if (x1 != -1 && x2 != -1 && x3 != -1 && x4 != -1){
   *result = (uint)((x1<<12)|(x2<<8)|(x3<<4)|x4);
   return s+4;
  ...// several } braces
 } 
 return 0;
}
void parse(nt_array_ptr<const char> s,
            array_ptr<uint> p : count(n), 
            int n) {
 array_ptr<uint> q : bounds(p,p+n) = p;
 while (s && q < p+n) {
   array_ptr<uint> r : count(1) =
     dyn_bounds_cast<array_ptr<uint>>(q,count(1));
   s = parse_utf16_hex(s,r);
   q++;
 }
}
  \end{lstlisting}

}

\caption{Parsing a String of UTF16 Hex Characters in \checkedc}
\label{fig:checkedc-example}
\end{figure}



\subsection{Checked Pointer Types}
\checkedc introduces three varieties of \emph{checked pointer}:
\begin{itemize}
\item \code{ptr<}$T$\code{>} types a pointer that is either null or
  points to a single object of type $T$.
\item \code{array_ptr<}$T$\code{>} types a pointer that is either null
  or points to an array of $T$ objects. The array width is defined
  by a \emph{bounds} expression, discussed below.
\item \code{nt_array_ptr<}$T$\code{>} is like
  \code{array_ptr<}$T$\code{>} except that the bounds expression
  defines the \emph{minimum} array width---additional objects may
  be available past the upper bound, up to a null terminator.
\end{itemize}
A bounds expression used with the latter two pointer types has three
forms:
\begin{itemize}
\item \code{count(}$e$\code{)} where $e$ defines the array's
  length. Thus, if pointer $p$ has bounds \code{count(n)} then the
  accessible memory is in the range $[p,p+$\code{n}$]$. Bounds
  expression $e$ must be side-effect free and may only refer to
  variables whose addresses are not taken, or adjacent \code{struct}
  fields.
\item \code{byte_count(}$e$\code{)} is like \code{count}, but
  expresses arithmetic using bytes, no objects; i.e.,
  \code{count(}$e$\code{)} used for \code{array_ptr<}$T$\code{>} is
  equivalent to \code{byte_count(}$e\times\texttt{sizeof}(T)$\code{)}
\item \code{bounds(}$e_l$,$e_h$\code{)} where $e_l$ and $e_h$ are
  pointers that bound the accessible region $[e_l,e_h)$ (the
  expressions are similarly restricted). Bounds
  \code{count(}$e$\code{)} is shorthand for
  \code{bounds(}$p, p + e$\code{)}. This most general form of bounds
  expression is useful for supporting pointer arithmetic.
\end{itemize}
  Dropping the bounds expression on an \code{nt_array_ptr} is equivalent
  to the bounds being \code{count(0)}.

The \checkedc compiler will instrument loads and stores of checked
pointers to confirm the pointer is non-null, and the access is within
the specified bounds. For pointers $p$ of type
\code{nt_array_ptr<}$T$\code{>}, such a check could spuriously fail if
the index is past $p$'s specified upper bound, but before the null
terminator. To address this problem, Checked C supports \emph{bounds
  widening}.
If $p$'s bounds expression is \code{bounds}$(e_l$,$e_h)$ a program may read from (but not
write to) $e_h$; when the compiler notices that a non-null character
is read at the upper bound, it will extend that bound to $e_h+1$.

\subsection{Example}
\label{sec:example}

Fig.~\ref{fig:checkedc-example} gives an example \checkedc
program.\footnote{Ported from the Parson JSON
parser, \url{https://github.com/kgabis/parson}} 
The function \code{parse_utf16_hex} on lines 1-15 takes a 
null-terminated pointer \code{s} as its argument, from which it attempts to read four
characters. These are interpreted as hex digits and converted to an
\code{uint} returned via parameter \code{result}. At first,
\code{s}~has no specific bounds annotation, which we can interpret as
\code{count(0)}; this means that \code{s[0]} may be read on line
5. The true branch of the conditional (which extends all the way to
the brace on line 13) is thus type-checked with \code{s} given a
\emph{widened} bound of \code{count(1)}. Likewise, the conditionals on
lines 6-8 each widen it one further; the widened pointer
(\code{s+4}) is returned on success.

The \code{parse} function on lines 16-26 repeatedly invokes
\code{parse_utf16_hex} with its parameter \code{s}, and fills out
array \code{p} whose declared length is the parameter \code{n}. Writes
happen via pointer \code{q}, which is updated using pointer
arithmetic. We specify its bounds as
\code{bounds(p,p+n)} to support this: even as \code{q} changes, variables \code{p} and \code{n} (and therefore also \code{q}'s bounds) do not. Converting from an
\code{array_ptr<uint>} to a \code{ptr<uint>}, done for the call on
line 23, requires proving the array has size at least 1. While this is true
because of the loop condition \code{q < p+n}, which is \code{q}'s
upper bound,  the compiler is not smart enough to figure this
out. To convince it, we manually insert a \emph{dynamic cast} via
\code{dyn_bounds_cast}, which is trusted at compile-time but confirmed
with a dynamic check at run-time.

While bounds checks are \emph{conceptually} inserted on every array
load and store, many of these are eliminated by LLVM\@. For example,
all of the pointer accesses to \code{s} on lines 5-8 are proved safe
at compile-time, so no bounds checks are inserted for
them. \citet{Elliott2018}
reported average run-time overheads of 8.6\% on a pointer-intensive
benchmark suite (49.3\% in one case); \citet{duanrefactoring} measured
no overhead at all on a port of FreeBSD's UDP and IP stacks to \checkedc.


\subsection{Other features}

\checkedc has other features not modeled in this paper. Two in regular
use are \emph{interop types}, which ascribe checked pointer types to
unported legacy code, notably in libraries; and \emph{generic types}
on both functions and \code{struct}s, for type-safe polymorphism. More
details about these can be found in the language specification.

\subsection{Spatial Safety and Backward Compatibility}
\checkedc is backward compatible with legacy C in the sense that all
legacy code will type-check and compile. However, only code that
appears in \emph{checked regions}, which we call \emph{checked code},
is spatially safe. Checked regions can be designated at the level of
files, functions, or individual code blocks, the first with a
\code{#pragma} and the latter two using the \code{checked}
keyword.\footnote{You can also designate \emph{unchecked} regions
  within checked ones.}  Within checked regions, both
legacy pointers and certain unsafe idioms (e.g., \emph{variadic} function
calls) are disallowed. The code in Fig.~\ref{fig:checkedc-example}
satisfies these conditions, and will type-check in a checked region.

How should we think about code that contains both checked and legacy
components? \citet{ruef18checkedc-incr} proved, for a simple
formalization of \checkedc, that \emph{checked code cannot be blamed}:
Any spatial safety violation is caused by the execution of unchecked
code. In this paper we extend that result to a richer formalization of
\checkedc.

%% file: formal.tex
\section{Formalization}\label{sec:formal}












\begin{figure}
  \small \centering
  $\begin{array}{l}
\begin{array}{lll}
\text{Function names:}~f&
       \text{Variables:}~ x
& \text{Integers:}~n::=\mathbb{Z} 
\end{array}
\\[0.5em]

\begin{array}{llcllcl}

\text{Mode:} & m & ::= & \cmode \mid \umode \\[0.5em]

\text{Bound:} & b & ::= & n \mid x \plus n \\
              & \bvar & ::= & (b,b) \\[0.5em]
  
     \text{Word Type:}& \tau &::=& \tint\mid \tptr{\omega}{m}
\\[0.5em]

\text{Type Flag:}&\kappa &::=& nt \mid \cdot
\\[0.5em]

\text{Type:}&\omega &::=& \tau \mid \tallarrayb{\bvar}{\tau}
\\[0.5em]

\text{Expression:}& e & ::= & 
\evalue{n}{\tau} \mid x \mid \emalloc{\omega} \mid\elet{x}{e}{e} \\[0.2em]
&&\mid&
\ecast{\tau}{e} \mid \edyncast{\tau}{e}\mid \ecall{f}{\overline{e}} \mid \estrlen{x} \\[0.2em]
&&\mid&
\ebinop{e}{e} \mid\estar{e}\mid\eassign{e}{e}\mid\eunchecked{e}
\\[0.2em]
&&\mid&\eif{e}{e}{e}
\end{array}
    \end{array}
  $
  \caption{\lang Syntax}
  \label{fig:checkc-syn}
\end{figure}



This section describes our formal model of \checkedc, called
\lang, making precise its syntax, semantics, and type system. It also
develops \lang's meta-theory, including type soundness and the blame
theorem.

\subsection{Syntax}\label{sec:syntax}

The syntax of \lang is given by the expression-based
language presented in Fig.~\ref{fig:checkc-syn}.

There are two notions of type in \lang.  Types $\tau$ classify
word-sized values including integers and pointers, while types
$\omega$ classify multi-word values such as arrays, null-terminated
arrays, and single-word-size values.
Pointer types ($\tptr{\omega}{m}$) include a mode annotation ($m$)
which is either checked (\cmode) or unchecked (\umode) and a type
($\omega$) denoting valid values that can be pointed to. Array types include both the type of
elements ($\tau$) and a bound ($\bvar$) comprised of an upper and
lower bound on the size of the array ($(b_l,b_h)$). Bounds $b$ are
limited to integer literals $n$ and expressions $x + n$.
Whether an array pointer is null terminated or not is determined by annotation
$\kappa$, which is $nt$ for null-terminated arrays, and $\cdot$
otherwise (we elide the $\cdot$ when writing the type). Here is the
corresponding \checkedc syntax for these types:
\[
\begin{array}{rcl}
$\code{array_ptr<}$\tau$\code{> : count(}$n$\code{)}$
&\Leftrightarrow& \tarrayptr{0}{n}{\tau}{\cmode}
\\[0.2em]
$\code{nt_array_ptr<}$\tau$\code{> : count(}$n$\code{)}$
&\Leftrightarrow& \tntarrayptr{0}{n}{\tau}{\cmode}
\end{array}
\]
As a convention we write $\tptr{\tarrayb{b}{\tau}}{\cmode}$ to mean
$\tptr{\tarray{0}{b}{\tau}}{\cmode}$, so the above examples could
be rewritten $\tptr{\tarrayb{n}{\tau}}{\cmode}$ and
$\tptr{\tntarrayb{n}{\tau}}{\cmode}$, respectively.

\lang expressions include literals ($n\!:\!\tau$), variables ($x$), memory
allocation ($\emalloc{\omega}$), let binding ($\elet{x}{e_1}{e_2}$),
static casts ($\ecast{\tau}{e}$), dynamic casts ($\edyncast{\tau}{e}$)
(assumed at compile-time and verified at run-time, see Sec.~\ref{sec:example}),
function calls ($\ecall{f}{\overline{e}}$), addition
($\ebinop{e_1}{e_2}$), pointer dereference and assignment ($\estar{e}$)
and ($\eassign{e_1}{e_2}$), resp.), unchecked blocks ($\eunchecked{e}$),
the \texttt{strlen} operation ($\estrlen{x}$), and conditionals ($\eif{e}{e_1}{e_2}$).

Integer literals $n$ are annotated with a type $\tau$ which can be either
$\tint$, or $\tptr{\omega}{m}$ in the case $n$ is being used as
a heap address (this is useful for the semantics);
$\evalue{0}{\tptr{\omega}{m}}$ (for any $m$ and $\omega$) represents
the $\enull$ pointer, as usual. 
The
$\texttt{strlen}$ expression operates on variables $x$
rather than arbitrary expressions to simplify managing
bounds information in the type system; the more general case can be
encoded with a \code{let}. We use a less verbose syntax for dynamic bounds
casts; e.g., the following %

\code{dyn_bounds_cast<array_ptr<}$\tau$\code{>>(}$e$\code{, count(}$n$\code{))}

\noindent
becomes $\edyncast{\tptr{\tarrayb{n}{\tau}}{\cmode}}{e}$. 

\lang aims to be simple enough to work with, but powerful enough to
encode realistic \checkedc idioms. For example, mutable local
variables can be encoded as immutable locals that point to the heap;
the use of \code{&} can be simulated with \code{malloc};
and loops can be encoded as recursive function calls. \code{struct}s are
not in Fig.~\ref{fig:checkc-syn} for space reasons, but they are
actually in our model, and developed in
\iftr
Appendix~\ref{appx:struct}.
\else
the supplemental report~\cite{checkedc-tech-report}.
\fi
C-style \code{union}s have no safe typing
in \checkedc, so we omit them. By default, functions are assumed to
be within checked regions; placing the body in an \code{unchecked}
expression relaxes this, and within that, checked regions can be
nested via function calls. Bounds are restricted slightly:
rather than allowing arbitrary sub-expressions, bounds must be either
integer literals or variables plus an integer offset, which accounts
for most uses of \code{bounds} in \checkedc programs. \lang bounds are
defined as relative offsets, not absolute ones, as in the second part
of Fig.~\ref{fig:checkedc-example}. We see no technical problem to
modeling absolute bounds, but it would be a pervasive change so we
have not done so.

  We have mechanized two models of \lang, one in Coq
  and one in PLT Redex~\cite{pltredex}, which is a semantic
  engineering framework implemented in Racket. Redex provides direct
  support for specifying the operational semantics and typing with logical
  rules, but then automatically makes them executable and subject to
  randomized testing, which is very useful during development.
  The model we present in the paper faithfully
  represents both mechanizations, but there are some differences for
  presentation purposes. For example, the paper and the Coq model use an explicit
  stack, whereas the Redex model uses \code{let} bindings to simulate
  one (simplifying term generation for randomized
  testing).
\iftr
  Appendix~\ref{app:model-diffs} 
\else
  The supplemental report~\cite{checkedc-tech-report}
\fi
  outlines the differences
  between the two models and the paper formalism.


\begin{figure}
{\small
$    \begin{array}{l}
    \begin{array}{lll}
\mu & ::= & \evalue{n}{\tau} \mid \bot\\
e & ::= & \ldots \mid \ret{x}{\mu}{e}\\
r & ::= & e \mid \enull \mid \ebounds\\
E &::=& \Box \mid \elet{x}{E}{e} \mid \ecall{f}{\overline{E}}\mid\ecast{\tau}{E} \mid \edyncast{\tau}{E} \\[0.2em]
&&\mid \ret{x}{\evalue{n}{\tau}}{E}\mid\ebinop{E}{e} \mid \ebinop{\evalue{n}{\tau}}{E}\mid\estar{E}\mid\eassign{E}{e} \\[0.2em]
&&\mid\eassign{\evalue{n}{\tau}}{E}\mid\eunchecked{E}\mid\eif{E}{e}{e}\\[0.2em]
\overline{E} &::=& E \mid  \evalue{n}{\tau},\overline{E} \mid \overline{E}, e
\end{array}
\\ \\
    \end{array} 
$
  \begin{mathpar}
    \inferrule{ m=\mode(E) \\
      e=E[e'] \\
      (\varphi,\heap,e') \longrightarrow (\varphi',\heap',e'')}
    {(\varphi,\heap,e)\longrightarrow_{m} (\varphi',\heap',E[e''])}

    \inferrule{ m=\mode(E) \\
      e=E[\eif{\estar{x}}{e_1}{e_2}] \\
      (\varphi,\heap,\eif{\estar{x}}{e_1}{e_2})
              \longrightarrow (\varphi',\heap',e')}
    {(\varphi,\heap,e)\longrightarrow_{m} (\varphi',\heap',E[e'])}
    \;\;\;{[\emph{prefer}]}

  \end{mathpar}
}
  \caption{\lang Semantics: Evaluation}
  \label{fig:c-context}
\end{figure}

\subsection{Semantics}\label{sec:semantics}



The operational semantics for \lang is defined as a small-step
transition relation with the judgment $ (\varphi,\heap,e)
\longrightarrow_m (\varphi',\heap',r)$. Here, $\varphi$ is a
\emph{stack} mapping from variables to values $\evalue{n}{\tau}$ and
$\heap$ is a \emph{heap} mapping addresses (integer literals) to
values $\evalue{n}{\tau}$; for both we ensure
$\fv(\tau)=\emptyset$.
While heap bindings can change, stack bindings are immutable---once
variable $x$ is bound to $\evalue{n}{\tau}$ in $\varphi$, that binding will not
be updated; we can model mutable stack variables as pointers into the
mutable heap.
As mentioned, value $\evalue{0}{\tau}$
represents a $\enull$ pointer when $\tau$ is a pointer type;
correspondingly, $\heap(0)$ should always be undefined.
The relation steps to a \emph{result} $r$,
which is either an expression or a $\enull$ or $\ebounds$ failure,
representing a null-pointer dereference or out-of-bounds access,
respectively. Such failures are a \emph{good} outcome; stuck states
(non-value expressions that cannot transition to a result $r$)
characterize undefined behavior.
The mode $m$ indicates whether the
stepped redex within $e$ was in a checked ($\cmode$) or
unchecked ($\umode$) region.

The rules for the main operational semantics
judgment---\emph{evaluation}---are given at the bottom of
Fig.~\ref{fig:c-context}.
The first rule takes an expression $e$, decomposes
it into an \emph{evaluation context} $E$ and a sub-expression $e'$
(such that replacing the hole $\Box$ in $E$ with $e'$ would yield
$e$), and then evaluates $e'$ according to the \emph{computation}
  relation $(\varphi,\heap,e') \longrightarrow (\varphi,\heap,e'')$,
whose rules are given in Fig.~\ref{fig:semantics}, discussed
shortly. 
The second rule handles conditionals $\eif{\estar{x}}{e_2}{e_3}$ in
redex position specially, delegating directly to the \textsc{S-IfNTT} computation
rule, which supports bounds widening; we discuss this rule
shortly. When the second and first rules could both apply, we always
prefer the second.\footnote{This approach is that of the PLT Redex model of \lang; the Coq
development uses a slightly simpler syntax to achieve the same
effect.}
The $\mode$ function
determines the mode when evaluating $e'$ based on the context $E$:
if the $\Box$ occurs within $(\eunchecked{E'})$ inside $E$, then the mode is
$\umode$; otherwise, it is $\cmode$. Evaluation contexts $E$ define a
standard left-to-right evaluation order. (We explain the
$\ret{x}{\mu}{e}$ syntax shortly.)

\begin{DIFnomarkup}
\begin{figure*}[t]
{\small
  \begin{mathpar}
        \inferrule[S-Def]{\heap(n)=\evalue{n_a}{\tau_a} }
    {(\varphi,\heap,\estar{\evalue{n}{\tptr{\tau}{m}}}) \longrightarrow (\varphi,\heap,\evalue{n_a}{\tau})}

    \inferrule[S-DefNull]{}{(\varphi,\heap,\estar{\evalue{0}{\tptr{\omega}{\cmode}}}) \longrightarrow (\varphi,\heap,\enull)}

        \inferrule[S-DefNTArray]{\heap(n)=\evalue{n_a}{\tau_a} \\ 0 \in [n_l,n_h]}
    {(\varphi,\heap,\estar{\evalue{n}{\tntarrayptr{n_l}{n_h}{\tau}{\cmode}}}) \longrightarrow (\varphi,\heap,\evalue{n_a}{\tau})}

    \inferrule[S-AssignArr]{\heap(n)=\evalue{n_a}{\tau_a}\\ 0 \in [n_l,n_h) }
      {(\varphi,\heap,\eassign{\evalue{n}{\tallarrayptr{n_l}{n_h}{\tau}{\cmode}}}{\evalue{n_1}{\tau_1}}) \longrightarrow (\varphi,\heap[n \mapsto \evalue{n_1}{\tau}],\evalue{n_1}{\tau})}

    \inferrule[S-Cast]
              {}
              {(\varphi,\heap,\ecast{\tau}{\evalue{n}{\tau'}}) \longrightarrow (\varphi,\heap,\evalue{n}{\varphi(\tau)})}

  \inferrule[S-DynCast]{
\varphi(\tptr{\tallarrayb{\bvar}{\tau}}{m})=\tallarrayptr{n_l'}{n_h'}{\tau_b}{m} \\ n_l' \le n_l\ \\ n_h \le n_h'}
    { (\varphi,\heap,\edyncast{\tptr{\tallarrayb{\bvar}{\tau}}{m}}{\evalue{n}{\tallarrayptr{n_l}{n_h}{\tau_a}{m}}}) \longrightarrow (\varphi',\heap',\evalue{n}{\tallarrayptr{n_l'}{n_h'}{\tau_b}{m}})}

 \inferrule[S-DynCastBound]{\varphi(\tptr{\tallarrayb{\bvar}{\tau}}{\cmode})=\tallarrayptr{n_l'}{n_h'}{\tau_b}{\cmode} \\ n_l' > n_l \vee n_h > n_h'}{ (\varphi,\heap,\edyncast{\tptr{\tallarrayb{\bvar}{\tau}}{\cmode}}{\evalue{n}{\tallarrayptr{n_l}{n_h}{\tau_a}{\cmode}}}) 
 \longrightarrow (\varphi',\heap',\ebounds)}

        \inferrule[S-Let]{}{(\varphi,\heap,\elet{x}{\evalue{n}{\tau}}{e}) \longrightarrow (\varphi[x\mapsto \evalue{n}{\tau}],\heap,\ret{x}{\varphi(x)}{e})}

    \inferrule[S-Ret]{}{(\varphi,\heap,\ret{x}{\mu}{\evalue{n}{\tau}}) \longrightarrow (\varphi[x\mapsto \mu],\heap,\evalue{n}{\tau})}

    \inferrule[S-Fun]{ \Xi(f) = \tau\;(\evalue{\overline{x}}{\overline{\tau}})\;e}
        {(\varphi,\heap,\ecall{f}{{\evalue{\overline{n}}{\overline{\tau_a}}}}) \longrightarrow
   (\varphi,\heap, \mathtt{let}\;\overline{x}={\evalue{\overline{n}}{(\overline{\tau}[\overline{n} / \overline{x}])}}\;\mathtt{in}\;\ecast{\tau[\overline{n} / \overline{x}]}{e})}

      \inferrule[S-StrWiden]{
         \varphi(x) = \evalue{n}{\tarrayptr{n_l}{n_h}{\tau}{\cmode}}\\ 0 \in [n_l,n_h] \\ n_a > n_h
\\ \heap(n+n_a) = 0 
\\\\ (\forall i. n \le i < n+n_a \Rightarrow (\exists n_i\;t_i. \heap(n+i) = \evalue{n_i}{\tau_i} \wedge n_i \neq 0))}
    {(\varphi,\heap,\estrlen{x}) \longrightarrow (\varphi[x\mapsto \evalue{n}{\tarrayptr{n_l}{n_a}{\tau}{\cmode}}],\heap,\evalue{n_a}{\tint})}

     \inferrule[S-IfNTT]{\varphi(x)=\evalue{n}{\tntarrayptr{n_l}{0}{\tau}{\cmode}} \\ \heap(n)\neq 0}
    {(\varphi,\heap,\eif{*x}{e_1}{e_2}) \longrightarrow (\varphi[x\mapsto \evalue{n}{\tntarrayptr{n_l}{1}{\tau}{\cmode}}],\heap,e_1)}


\end{mathpar}
}
\caption{\lang Semantics: Computation (Selected Rules)}
\label{fig:semantics}
\end{figure*}
\end{DIFnomarkup}



\begin{figure}[t]
{\small
{\captionsetup[lstlisting]{margin = 8 mm}
  \begin{lstlisting}[xleftmargin=8 mm]
nt_array_ptr<char> safe_strcat
   (nt_array_ptr<char> dst : count(n),
    nt_array_ptr<char> src : count(0), int n) {
  int x = strlen(dst);
  int y = strlen(src);
  nt_array_ptr<char> c : count(n) =
    dyn_bounds_cast
           <nt_array_ptr<char>>(dst,count(n));
    // sets c == dst with bound n (not x)
  if (x+y < n) {
    for (int i = 0; i < y; ++i)
      *(c+x+i) = *(src+i);
    *(c+x+y) = '\0';
    return dst;
  }
  return null;
}
  \end{lstlisting}
}
}
\caption{Implementation of safe \code{strcat}}
\label{fig:strcat-ex}
\end{figure}

Fig.~\ref{fig:semantics} shows selected rules for the computation
relation; we explain them with the help of the example in
Fig.~\ref{fig:strcat-ex},
  which defines a 
  safe version of \code{strcat} (using actual Checked C syntax).  The
  function takes a target 
  pointer \code{dst} of capacity \code{n}, where the first null
  character (determined by \code{strlen}) is at index \code{x} where
  $0 \leq $\code{x}$ \leq n$. It concatenates the \code{src} buffer to
  the end of \code{dst} as long as \code{dst} has sufficient space.


\myparagraph{Pointer accesses}
The rules for dereference and assignment operations---\textsc{S-Def},
\textsc{S-DefNull}, \textsc{S-DefNTArray}, and
\textsc{S-AssignArr}---illustrate how the semantics checks bounds.
Rule \textsc{S-DefNull} transitions attempted null-pointer
dereferences to $\enull$, whereas \textsc{S-Def} dereferences a
non-null (single) pointer.
  When $\enull$ is returned by the
computation relation, the evaluation relation halts the entire
evaluation with $\enull$ (using a rule not shown in Fig.~\ref{fig:c-context}); it
does likewise when $\ebounds$ is returned (see below).


\textsc{S-AssignArr} assigns to an array as long as 0 (the point of
dereference) is within the bounds designated by the pointer's annotation
and strictly less than the upper bound. For the assignment rule,
arrays are treated uniformly whether they are null-terminated or not
($\kappa$ can be $\cdot$ or $nt$)---the semantics does not search past
the current position for a null terminator. The program
can widen the bounds as needed, if they currently precede the null
terminator: \textsc{S-DefNTArray}, which dereferences an NT array
pointer, allows an upper bound of $0$, since the program may read, but
not write, the null terminator. A separate rule (not shown) handles
normal arrays.

\myparagraph{Casts}
Static casts of a literal $n\!:\!\tau'$ to a type $\tau$ are handled
by \textsc{S-Cast}. In a type-correct program, such casts are
confirmed safe by the type system. To evaluate a cast, the rule
updates the type annotation on $n$. Before doing so, it must
``evaluate'' any variables that occur in $\tau$ according to their
bindings in $\varphi$. For example, if $\tau$ was
$\tarrayptr{0}{x+3}{\tint}{\cmode}$, then $\varphi(\tau)$ would
produce $\tarrayptr{0}{5}{\tint}{\cmode}$ if $\varphi(x) = 2$.

Dynamic casts are accounted for by \textsc{S-DynCast} and
\textsc{S-DynCastBound}. In a type-correct program, such casts are
assumed correct by the type system, and later confirmed by the
semantics. As such, a dynamic cast will cause a $\ebounds$ failure if
the cast-to type is incompatible with the type of the target pointer,
as per the $n_l' > n_l \vee n_h > n_h'$ condition in
\textsc{S-DynCastBound}. An example use of dynamic casts is given on
line~7 in Fig.~\ref{fig:strcat-ex}. The values of \code{x} and
\code{n} might not be known statically, so the type system cannot
confirm that \code{x <= n}; the dynamic cast assumes this inequality
holds, but then checks it at run-time.

\myparagraph{Binding and Function Calls}
The semantics handles variable scopes using the special $\erettext$
form. \textsc{S-Let} evaluates to a configuration whose stack
is $\varphi$ extended with a binding for $x$, and whose expression is
$\ret{x}{\varphi(x)}{e})$ which remembers $x$ was previously bound to
$\varphi(x)$; if it had no previous binding, $\varphi(x) =
\bot$. Evaluation proceeds on $e$ until it becomes a literal
$n\!:\!\tau$, in which case \textsc{S-Ret} restores the saved
binding (or $\bot$) in the new stack, and evaluates to
$n\!:\!\tau$. 

Function calls are handled by \textsc{S-Fun}. Recall that array
bounds in types may refer to in-scope variables; e.g., parameter
\code{dst}'s bound \code{count(n)} refers to parameter \code{n} on lines
2-3 in Fig.~\ref{fig:strcat-ex}. A call to function $f$ causes $f$'s
definition to be retrieved from $\Xi$,
which maps function names to
forms $\tau\;(\evalue{\overline{x}}{\overline{\tau}})\;e$, where
$\tau$ is the return type, $(\evalue{\overline{x}}{\overline{\tau}})$
is the parameter list of variables and their types, and $e$ is the
function body. The call is expanded into a \texttt{let} which binds
parameter variables $\overline{x}$ to the actual arguments
$\overline{n}$, but annotated with the parameter types
$\overline{\tau}$ (this will be safe for type-correct programs). The
function body $e$ is wrapped in a static cast
$(\tau[\overline{n} / \overline{x}])$ which is the function's return
type but with any parameter variables $\overline{x}$ appearing in that
type substituted with the call's actual arguments $\overline{n}$. To
see why this is needed, suppose that \code{safe_strcat} in
Fig.~\ref{fig:strcat-ex} is defined to return a
\code{nt_array_ptr<int>:count(n)} typed term, and assume that we
perform a \code{safe_strcat} function call as
\code{x=safe_strcat(a,b,10)}. After the evaluation of \code{safe_strcat}, the
function returns a value with type \code{nt_array_ptr<int>:count(10)}
because we substitute bound variable \code{n} in the 
defined return type with \code{10} from the function call's
argument list.
Note that the \textsc{S-Fun} rule replaces the
  annotations $\overline{\tau_a}$ with
  $\overline{\tau}$ (after instantiation) from the function's
  signature. Using $\overline{\tau_a}$ when executing the body of
the function has no impact on the soundness of \lang, but will violate
Theorem~\ref{simulation-thm}, which we introduce in Sec.~\ref{sec:compilation}.

\myparagraph{Bounds Widening}
Bounds widening occurs when branching on a dereference of a NT array
pointer, or when performing $\estrlentext$. The latter is most useful
when assigned to a local variable so that subsequent code can use the
result, e.g., $e$ in $\elet{x}{\estrlen{y}}{e}$.  Lines 4 and 5 in
Fig.~\ref{fig:strcat-ex} are examples. 
The widened upper bound
precipitated by $\estrlen{y}$ is extended beyond the lifetime of $x$, as long as $y$ is live.
For example, \code{x}'s scope in line 4 at runtime is
the whole function body in \code{safe_strcat} because the lifetime of the pointer \code{dst} is in the function body.
This is different from the \checkedc specification, 
which only allows bound widening to happen within the scope of \code{x}, and restoring old bound values once \code{x} dies.
We allow widening to persist outside the scope at run-time as long as
we are within the stack frame, and we show
this does not necessarily require the use of fat pointers in
Sec.~\ref{sec:compilation}.

Rule \textsc{S-StrWiden} implements $\estrlentext$ widening. The
predicate
$\forall i. n \le i < n+n_a \Rightarrow (\exists n_i\;t_i. \heap(n+i)
= \evalue{n_i}{\tau_i} \wedge n_i \neq 0))$ aims to find a position
$n+n_a$ in the NT array that stores a null character, where no
character as indexes between $n$ and $n+n_a$ contains one. (This rule
handles the case when $n_a > n_h$, the $n_a \le n_h$ case is handled
by a normal $\estrlentext$ rule; see 
\iftr
Appendix~\ref{sec:rem-semantics}.)
\else
the supplemental report~\cite{checkedc-tech-report}.)
\fi

Rule \textsc{S-IfNTT} performs bounds widening on $x$ when the
dereference $\estar{x}$ is not at the null terminator, but the
pointer's upper bound is 0 (i.e., it's at the end of its known
range). $x$'s upper bound is incremented to 1, and this count persists
as long as $x$ is live.  For example, \code{s}'s increment (lines
5--8) is live until the return of the function in
Fig.~\ref{fig:checkedc-example}; thus, line 11 is valid because
\code{s}'s upper bound is properly extended. 

\subsection{Typing}\label{sec:type-system}


We now turn to the \lang type system.
%
The typing judgment has the form $\Gamma;\Theta\vdash_m e : \tau$,
which states that in a type environment $\Gamma$ (mapping variables to
their types) and a predicate environment $\Theta$ (mapping integer-typed
variables to Boolean predicates), expression $e$ will have type $\tau$ if evaluated
in mode $m$. Key rules for this judgment are given in
Fig.~\ref{fig:type-system-1}. In the rules, $m \le m'$ uses
the two-point lattice with $\umode < \cmode$.
All remaining rules
are given in
\iftr
Appendix~\ref{sec:literal-pointer-typing}~and~\ref{rem-type}.
\else
the supplemental report~\cite{checkedc-tech-report}.
\fi

\begin{DIFnomarkup}
\begin{figure*}[t]
{\small
  \begin{mathpar}
    \inferrule[T-DefArr]
              {m \leq m' \\\\ \Gamma;\Theta \vdash_{m} e : \tptr{\tallarrayb{\bvar}{\tau}}{m'}}
              {\Gamma;\Theta \vdash_m \estar{e} : \tau}

    \inferrule[T-AssignArr]
              {\Gamma; \Theta \vdash_m e_1 : \tptr{\tallarrayb{\bvar}{\tau}}{m'}\\\\
                \Gamma; \Theta \vdash_m e_2 : \tau' \\
                \tau'\sqsubseteq \tau\\
                m \leq m'}
              {\Gamma; \Theta \vdash_m \eassign{e_1}{e_2} : \tau}

    \inferrule[T-Unchecked]
              {\Gamma;\Theta \vdash_u e : \tau}
              {\Gamma;\Theta \vdash_m \eunchecked{e} : \tau}

    \inferrule[T-Cast]
              {m = \cmode \Rightarrow \tau \neq\tptr{\tau''}{\cmode} \text{ for any $\tau''$} \\\\
                \Gamma;\Theta \vdash_m e : \tau'}
              {\Gamma;\Theta \vdash_m \ecast{\tau}{e} : \tau}
              
     \inferrule[T-CastCheckedPtr]
               {\Gamma;\Theta \vdash_m e : \tau' \\
                 \tau' \sqsubseteq \tptr{\tau}{\cmode}}
               {\Gamma;\Theta \vdash_m \ecast{\tptr{\tau}{\cmode}}{e} : \tptr{\tau}{\cmode}}
                
      \inferrule[T-DynCast]
                {\Gamma;\Theta \vdash_m e : \tptr{\tallarrayb{\bvar'}{\tau}}{m}}
                {\Gamma;\Theta \vdash_m \edyncast{\tptr{\tallarrayb{\bvar}{\tau}}{m}}{e} : \tptr{\tallarrayb{\bvar}{\tau}}{m}}
                
      \inferrule[T-If ]
        {\Gamma; \Theta \vdash_m e : \tau \\\\ \Gamma; \Theta \vdash_m e_1 : \tau_1 \\\\ \Gamma; \Theta \vdash_m e_2 : \tau_2}
        {\Gamma; \Theta \vdash_m \eif{e}{e_1}{e_2} : \tau_1 \sqcup \tau_2}
        
      \inferrule [T-IfNT]
                { \Gamma; \Theta \vdash_m x : \tptr{\tntarrayb{(b_l,0)}{\tau}}{c} \\\\
                   \Gamma[x \mapsto \tptr{\tntarrayb{(b_l,1)}{\tau}}{c}]; \Theta \vdash_m e_1 : \tau_1\\\\
                  \Gamma;\Theta \vdash_m e_2 : \tau_2 }
                {\Gamma;\Theta \vdash_m \eif{\estar{x}}{e_1}{e_2} : \tau_1 \sqcup \tau_2}

      \inferrule[T-Str]
                {\Gamma; \Theta \vdash_m e :  \tptr{\tntarrayb{\bvar}{\tau_a}}{m} 
                }
                {\Gamma;\Theta \vdash_m \estrlen{e} : \tint}

      \inferrule[T-LetStr]
                {\Gamma(y) = \tntarrayptr{b_l}{b_h}{\tau_a}{\cmode} \\ x \not\in \fv(\tau) \\\\
                 \Gamma[x\mapsto \tint,y\mapsto [\tntarrayptr{b_l}{x}{\tau_a}{\cmode}];\Theta[x\mapsto \texttt{ge}\_0] \vdash_m e : \tau}
                {\Gamma;\Theta \vdash_m \elet{x}{\estrlen{y}}{e} : \tau}

   \inferrule[T-Let]
    { x\in \fv(\tau') \Rightarrow e_1 \in \text{Bound} \\\\
        \Gamma;\Theta \vdash_m e_1 : \tau \\
           \Gamma[x\mapsto \tau];\Theta \vdash_m e_2 : \tau'
             }
    {\Gamma;\Theta \vdash_m \elet{x}{e_1}{e_2} : \tau'[e_1 / x]}

\inferrule[T-Fun]
    {\Xi(f) = \tau\;(\overline{x}:\overline{\tau})\;e \\
        \Gamma; \Theta \vdash_m \overline{e} : \overline{\tau'} \\
             \overline{\tau'} \sqsubseteq 
               \overline{\tau}[\overline{e} / \overline{x}]}
    {\Gamma; \Theta \vdash_m f(\overline{e}) : \tau[\overline{e} / \overline{x}]}

                \inferrule[T-Ret]
    {\Gamma(x)\neq \bot \\
          \Gamma;\Theta \vdash_m e : \tau}
    {\Gamma;\Theta \vdash_m \eret{x}{\mu}{e} : \tau}

  \end{mathpar}
}
\caption{Selected type rules}
\label{fig:type-system-1}
\end{figure*}
\end{DIFnomarkup}

\myparagraph{Pointer Access}
Rules \textsc{T-DefArr} and \textsc{T-AssignArr} type-check array
dereference and assignment operations resp., returning the type of
pointed-to objects; rules for pointers to single objects are
similar.
The condition $m\le m'$ ensures that unchecked pointers can only
be dereferenced in unchecked blocks;
the type rule for
$\eunchecked{e}$ sets $m = \umode$ when checking $e$.
The rules do not attempt to reason whether the access is in bounds;
this check is deferred to the semantics.


\myparagraph{Casting and Subtyping}
Rule \textsc{T-Cast} rule forbids casting to checked pointers when in checked
regions (when $m = \cmode$), but $\tau$ is unrestricted when
$m = \umode$. The \textsc{T-CastCheckedPtr} rule
permits casting from an expression of type $\tau'$ to a checked pointer when
$\tau' \sqsubseteq \tptr{\tau}{\cmode}$. This subtyping relation
$\sqsubseteq$ is given in Fig.~\ref{fig:checkc-subtype}; the many
rules ensure the relation is transitive. Most of the rules handle
casting between array pointer types. The second rule 
$0\le b_l \wedge b_h \le 1 \Rightarrow \tptr{\tau}{m}\sqsubseteq
\tarrayptr{b_l}{b_h}{\tau}{m}$ permits treating a singleton
pointer as an array pointer with $b_h\le 1$ and $0 \le b_l$.

Since bounds expressions may
contain variables, determining assumptions like $b_l \leq b_l'$
requires reasoning about those variables' possible values. The type
system uses $\Theta$ to make such reasoning more precise.\footnote{Technically, the subtyping relation $\sqsubseteq$ and the bounds
  ordering relation $\leq$ are parameterized by $\Theta$; this fact is
  implicit to avoid clutter.} $\Theta$ is a map from variables $x$ to
predicates $P$, which have the form $P ::= \top \;|\; \texttt{ge}\_0$.
If $\Theta$ maps $x$ to $\top$, that means that the variable can
possibly be any value; $\texttt{ge}\_0$ means that $x \ge 0$. We will
see how $\Theta$ gets populated and give a detailed example of
subtyping below.\footnote{As it turns out, the subtyping relation is
  also parameterized by $\varphi$, which is needed when type checking
  intermediate results to prove type preservation; source programs
  would always have $\varphi = \emptyset$. Details are in
\iftr
  Appendix~\ref{app:le}.
\else
  the supplemental report~\cite{checkedc-tech-report}.
\fi
}

Rule \textsc{T-DynCast} typechecks dynamic casting operations, which
apply to array pointer types only. The cast is accepted by the type
system, as its legality will be checked by the semantics.

\begin{DIFnomarkup}
\begin{figure}
{\small
\begin{center}
  $\begin{array}{r@{~}c@{~}l@{~}c@{~}l}
    &&\multicolumn{1}{r}{\tau} &\sqsubseteq& \tau\\[0.6em]
    0\le b_l \wedge b_h \le 1 &\Rightarrow& \tptr{\tau}{m}&\sqsubseteq& \tarrayptr{b_l}{b_h}{\tau}{m}\\[0.6em]
    b_l \le 0 \wedge 1 \le b_h &\Rightarrow& \tarrayptr{b_l}{b_h}{\tau}{m} &\sqsubseteq& \tptr{\tau}{m}\\[0.6em]
    b_l \le 0 \wedge 1 \le b_h &\Rightarrow& \tntarrayptr{b_l}{b_h}{\tau}{m} &\sqsubseteq& \tptr{\tau}{m}\\[0.6em]
    b_l \le b_l' \wedge b_h' \le b_h &\Rightarrow& \tntarrayptr{b_l}{b_h}{\tau}{m} &\sqsubseteq& \tarrayptr{b_l'}{b_h'}{\tau}{m}\\[0.6em]
    b_l \le b_l' \wedge b_h' \le b_h &\Rightarrow& \tallarrayptr{b_l}{b_h}{\tau}{m} &\sqsubseteq& \tallarrayptr{b_l'}{b_h'}{\tau}{m}
    \end{array}
  $
\end{center}
}
  \caption{Subtyping}
  \label{fig:checkc-subtype}
\end{figure}
\end{DIFnomarkup}

\myparagraph{Bounds Widening}
The bounds of NT array pointers may be widened at conditionals, and calls to $\estrlentext$.  Rule \textsc{T-If} handles normal
branching operations; rule \textsc{T-IfNT} is specialized to the case
of branching on $\estar{x}$ when $x$ is a NT array pointer whose upper
bound is 0. In this case, true-branch $e_1$ is checked with $x$'s type
updated so that its upper bound is incremented by 1; the else-branch
$e_2$ is type-checked under the existing assumptions. For both rules,
the resulting type is the join of the types of the two branches
(according to subtyping). This is important for the situation when $x$
itself is part of the result, since $x$ will have
different types in the two branches.


Rule \textsc{T-Str} handles the case for when $\estrlentext(y)$ does
not appear in a let binding. Rule \textsc{T-LetStr} handles the case
when it does, and performs bounds widening. The result of the call is
stored in variable $x$, and the type of $y$ is updated in $\Gamma$ when
checking the let-body $e$ to indicate that $x$ is $y$'s upper
bound. Notice that the lower bound $b_l$ is unaffected by the call to
$\estrlentext(y)$; this is sound because we know that $\estrlentext$
will always return a result $n$ such that $n \ge b_h$, the current
view of $x$'s upper bound.
The type rule tracks $\estrlentext$'s widened bounds
within the scope of $x$, while the bound-widening effect in the semantics applies to the lifetime of $y$.
Our type preservation theorem in Sec.~\ref{sec:theorem}
shows that our type system is a sound model of the \CoreChkC semantics,
and we discuss how we guarantee that the behavior of our compiler formalization and the semantics matches in Sec.~\ref{sec:compilation}.

This rule also extends $\Theta$ when checking $e$, adding a predicate
indicating that $x\ge 0$. To see how this information is used,
consider this example.  The \code{return} on line 14 of
Fig.~\ref{fig:strcat-ex} has an implicit static cast from the
returned expression to the declared function type (see rule
\textsc{T-Fun}, described below). In type checking the $\estrlentext$
on line 4, we insert a predicate in $\Theta$ showing
\code{x}$\;\ge 0$.  The static cast on line 14 is valid according to
the last line in Fig.~\ref{fig:checkc-subtype}:

{\small
$$
\tallarrayptr{0}{x}{\tau}{c} \sqsubseteq \tallarrayptr{0}{0}{\tau}{c}
$$
}
because $0 \leq 0$ and $0 \leq x$, where the latter holds since
$\Theta$ proves $x \ge 0$.  Without $\Theta$, we would need a
dynamic cast.

In our formal presentation, $\Theta$ is quite simple and is just meant
to illustrate how static information can be used to avoid dynamic
checks; it is easy to imagine richer environments of facts that can be
leveraged by, say, an SMT solver as part of the subtyping
check \cite{LiquidHaskell,Refinementlh}.

\myparagraph{Dependent Functions and Let Bindings}
Rule \textsc{T-Fun} is the standard dependent function call rule. It
looks up the definition of the function in the function environment
$\Xi$, type-checks the actual arguments $\overline{e}$ which have
types $\overline{\tau'}$, and then confirms that each of these types
is a subtype of the declared type of $f$'s corresponding parameter. Because
functions have dependent types, we substitute each parameter $e_i$ for
its corresponding parameter $x_i$ in both the parameter types and the
return type. Consider the \code{safe_strcat} function in
Fig.~\ref{fig:strcat-ex}; its parameter type for \code{dst} 
depends on \code{n}. The \textsc{T-Fun} rule will substitute 
\code{n} with the argument at a call-site.

Rule \textsc{T-Let} types a $\elettext$ expression, which also admits
type dependency. In particular, the result of evaluating a $\elettext$
may have a type that refers to one of its bound variables (e.g., if
the result is a checked pointer with a variable-defined bound); if so,
we must substitute away this variable once it goes out of scope. Note
that we restrict the expression $e_1$ to syntactically match the
structure of a Bounds expression $b$ (see Fig.~\ref{fig:checkc-syn}).

Rule \textsc{T-Ret} types a $\erettext$ expression, which does not
appear in source programs but is introduced by the semantics when
evaluating a let binding (rule \textsc{S-Let} in
Fig.~\ref{fig:semantics}); this rule is needed for the preservation
proof. After the evaluation of a let binding a variable $x$ concludes,
we need to restore any prior binding of $x$, which is either
$\bot$ (meaning that there is no $x$ originally) or some value
$\evalue{n}{\tau}$.

\subsection{Type Soundness and Blame}\label{sec:theorem}

In this subsection, we focus on our main meta-theoretic results about
\lang: type soundness (progress and preservation) and
blame.
  These proofs have been carried out in our
  Coq model, found at \url{https://github.com/plum-umd/checkedc}.

The type soundness theorems rely on several notions of
\emph{well-formedness}:

\begin{defi}[Type Environment Well-formedness]\label{type-wellformed}
A type environment $\Gamma$ is well-formed iff every variable mentioned as type bounds in $\Gamma$ are bounded by $\tnat$ typed variables in $\Gamma$.
\end{defi}

\begin{defi}[Heap Well-formedness]
A heap $\heap$ is well-formed iff (i) $\heap(0)$ is undefined, and
(ii) for all $\evalue{n}{\tau}$ in the range of $\heap$, type $\tau$
contains no free variables. 
\end{defi}

\begin{defi}[Stack Well-formedness]
A stack snapshot $\varphi$ is well-formed iff
for all $\evalue{n}{\tau}$ in the range of $\varphi$, type $\tau$
contains no free variables. 
\end{defi}

We also need to introduce a notion of
\emph{consistency}, relating heap environments before and after a
reduction step, and type environments, predicate sets, and stack
snapshots together.

\begin{defi}[Stack Consistency]
A type environment $\Gamma$, variable predicate set $\Theta$, and
stack snapshot $\varphi$ are consistent---written $\Gamma;\Theta\vdash
\varphi$---iff for every variable $x$, $\Theta(x)$ is defined implies
$\Gamma(x) = \tau$ for some $\tau$ and 
$\varphi(x) =\evalue{n}{\tau'}$ for some $n,\tau'$ where $\tau' \sqsubseteq \tau$. 
\end{defi}

\begin{defi}[Stack-Heap Consistency]
A stack snapshot $\varphi$ is consistent with heap $\heap$---written $\heap \vdash \varphi$---iff
for every variable $x$, $\varphi(x)= \evalue{n}{\tau}$ implies $\heap;\emptyset \vdash n:\tau$.
\end{defi}

\begin{defi}[Heap-Heap Consistency]
A heap $\heap'$ is consistent with $\heap$---written $\heap \triangleright \heap'$---iff
for every constant $n$, $\heap;\emptyset \vdash n:\tau$ implies $\heap';\emptyset \vdash n:\tau$.
\end{defi}

Moreover, as a program evaluates, its expression may contain literals
$\evalue{n}{\tau}$ where $\tau$ is a pointer type, i.e., $n$ is an
index in $\heap$ (perhaps because $n$ was chosen by
$\mathtt{malloc}$). The normal type-checking judgment for $e$ is
implicitly parameterized by $\heap$, and the rules for type-checking
literals confirm that pointed-to heap cells are compatible with
(subtypes of) the pointer's type annotation; in turn this check may
precipitate checking the type consistency of the heap itself. We
follow the same approach as \citet{ruef18checkedc-incr}, and show the
rules in
  Fig.~\ref{fig:const-type}; the judgment
  $\heap;\sigma \vdash n : \tau$ is used to confirm literal
  well-typing, where $\sigma$ is a set of pointer literals already
  checked in $\heap$ (to allow pointer cycles). See
  \iftr
  Appendix~\ref{sec:literal-pointer-typing}
  \else
  the supplemental report~\cite{checkedc-tech-report}
  \fi
   for further discussion.

Progress now states that terms that don't reduce are either values or their
mode is unchecked:

\begin{thm}[Progress]\label{thm:progress}

For any \checkedc program $e$, heap $\heap$, stack
$\varphi$, type environment $\Gamma$, and variable predicate set $\Theta$
that are all are well-formed, consistent
($\Gamma;\Theta\vdash \varphi$ and $\heap \vdash \varphi$) and well
typed ($\Gamma;\Theta\vdash_{\cmode} e : \tau$ for some $\tau$),
one of the following holds:

\begin{itemize}

\item $e$ is a value ($\evalue{n}{\tau}$).

\item there exists $\varphi'$ $\heap'$ $r$, such that $(\varphi,\heap,e) \longrightarrow_m (\varphi',\heap',r)$.

\item $m = \umode$, or there exists $E$ and $e'$, such that $e = E[e']$ and $\mode(E) = \umode$.

\end{itemize}
\end{thm}
{\em Proof:} By induction on the typing derivation.

\begin{DIFnomarkup}
 \begin{figure}[t]
 {\small
 \text{Type rules for constants and variables:}
 \begin{mathpar}
   \inferrule[T-Var]
       {x : \tau \in \Gamma}
       {\Gamma;\Theta \vdash_m x : \tau}
  
   \inferrule[T-Const]
       {\fv(\tau) = \emptyset \\ \heap;\emptyset \vdash n : \tau}
       {\Gamma;\Theta\vdash_m \evalue{n}{\tau} : \tau}
 \end{mathpar}
     
 \text{Rules for checking constant pointers:}
 \begin{mathpar}
   \inferrule
       {}
       {\heap;\sigma \vdash n : \tint}
  
   \inferrule
       {}
       {\heap;\sigma \vdash n : \tptr{\omega}{\umode}}
  
   \inferrule
       {}
       {\heap;\sigma \vdash 0 : \tptr{\omega}{\cmode}}
  
   \inferrule
       {(\evalue{n}{\tptr{\omega}{\cmode}})\in \sigma}
       {\heap;\sigma \vdash n : \tptr{\omega}{\cmode}}
  
   \inferrule
       {\forall i \in [0,\size(\omega)) .
            \heap;(\sigma \cup \{(n:\tptr{\omega}{\cmode})) \}\vdash \heap(n+i)}
       {\heap;\sigma \vdash n : \tptr{\omega}{\cmode}}
 \end{mathpar}
 }
 \caption{Type Rules for Constants/Variables}
 \label{fig:const-type}
 \end{figure}
\end{DIFnomarkup}

\noindent
Preservation states that a reduction step preserves both the
type and consistency of the program being reduced.

\begin{thm}[Preservation]
For any \checkedc program $e$, heap $\heap$, stack
$\varphi$, type environment $\Gamma$, and variable predicate set $\Theta$
that are all are well-formed, consistent
($\Gamma;\Theta\vdash \varphi$ and $\heap \vdash \varphi$) and well
typed ($\Gamma;\Theta\vdash_{\cmode} e : \tau$ for some $\tau$), if there exists $\varphi'$,
$\heap'$ and $e'$, such that $(\varphi,\heap,e)
\longrightarrow_{\cmode} (\varphi',\heap',e')$, then $\heap'$ is
consistent with $\heap$ ($\heap \triangleright \heap'$) and there exists
$\Gamma'$, $\Theta'$ and $\tau'$ that are well formed, consistent
($\Gamma';\Theta'\vdash \varphi'$ and $\heap' \vdash \varphi'$) and
well typed ($\Gamma';\Theta' \vdash_{\cmode} e: \tau'$), where
$\tau'\sqsubseteq \tau$.
\end{thm}
{\em Proof:} By induction on the typing derivation.
\smallskip

Using these two theorems we can prove our main result, {\em blame},
which states that
  if a well-typed program is \emph{stuck}---expression
$e$ is a non-value that cannot take a step\footnote{Note that
  $\ebounds$ and $\enull$ are \emph{not} stuck expressions---they represent a
  program terminated by a failed run-time check. A program that tries to access $\heap{n}$
  but $\heap$ is undefined at $n$ will be stuck, and violates spatial
  safety.}---the cause must be the
(past or imminent) execution of code in an unchecked region.


\begin{thm}[The Blame Theorem]\label{thm:blame} For any \checkedc
  program $e$, heap $\heap$, stack
$\varphi$, type environment $\Gamma$, and variable predicate set $\Theta$
that are well-formed and consistent
($\Gamma;\Theta\vdash \varphi$ and $\heap \vdash \varphi$),
if $e$ is well-typed ($\varphi;\Theta\vdash_{\cmode} e :
\tau$ for some $\tau$) and there exists
$\varphi_i$, $\heap_i$, $e_i$, and $m_i$ for $i\in [1,k]$, such that
$(\varphi,\heap,e) \longrightarrow_{m_1} (\varphi_1,\heap_1,e_1)\longrightarrow_{m_2} ...\longrightarrow_{m_k} (\varphi_k,\heap_k,r)$ and $r$ is \emph{stuck}, then
there exists $j \in [1,k]$, such that $m_j = \umode$, or there exists $E$ and $e'$, such that $r = E[e']$ and $\mode(E) = \umode$.
\end{thm}
{\em Proof:} By induction on the number of steps of the \checkedc
evaluation ($\longrightarrow_m^*$), using progress and preservation to
maintain the invariance of the assumptions.

  Compared to \citet{ruef18checkedc-incr}, proofs for
  \lang were made challenging by the addition of dependently typed
  functions and dynamic arrays, and the need to handle bounds widening for NT
  array pointers. These features required changes in the runtime
  semantics (adding a stack, and dynamically changing bounds) and in
  compile-time knowledge of them (to soundly typing widened bounds).

%% file: compilation.tex
\section{Compilation}\label{sec:compilation}

The semantics of \lang uses annotations on pointer literals in order
to keep track of array bounds information, which is used in premises
of rules like \textsc{S-DefArray} and
  \textsc{S-AssignArr} to prevent spatial safety violations. However, in the real
implementation of \checkedc, which extends Clang/LLVM, these annotations
are not present---pointers are represented as a single
machine word with no extra metadata, and bounds
  checks are not handled by the machine, but inserted by the
  compiler.

This section shows how \lang
annotations can be safely erased: using static information a compiler
can insert code to manage and check bounds metadata, with no loss of
expressiveness. We present a compilation algorithm that converts from
\lang to \elang, an untyped language without metadata
annotations. The syntax and semantics \elang
  closely mirrors that of \lang; it differs only in that literals lack
  type annotations and its operational rules perform no
  bounds and null checks, which are instead inserted during
  compilation. Our compilation algorithm is evidence that \lang's
  semantics, despite its apparent use of fat pointers, faithfully
  represents Checked C's intended behavior. The algorithm also sheds
  light on how compilation can be implemented in the real Checked C
  compiler, while eschewing many important details (\elang has many 
  differences with LLVM IR).

Compilation is defined by extending \lang's
typing judgment thusly:
\[\Gamma;\Theta;\rho \vdash_m e \gg \dot e:\tau\]
There is now a \elang output $\dot e$ and an input $\rho$, which maps
each \code{nt_array_ptr} variable \code{p} to a pair of \emph{shadow
  variables} that keep \code{p}'s up-to-date upper and lower bounds; these
may differ from the bounds in \code{p}'s type due to bounds
widening.\footnote{Since lower bounds are never widened, the
  lower-bound shadow variable is unnecessary; we include it for uniformity.}

We formalize rules for this judgment in PLT Redex~\cite{pltredex},
following and extending our Coq development for \lang. To give
confidence that compilation is correct, we use Redex's property-based
random testing support to show that compiled-to $\dot e $ simulates
$e$, for all $e$.


\subsection{Approach}

Due to space constraints, we explain the rules for compilation by
example, using a C-like syntax; the complete rules are given in
\iftr
Appendix~\ref{appx:comp1}.
\else
the supplemental report~\cite{checkedc-tech-report}.
\fi
Each rule performs up to three tasks: (a) conversion of $e$ to
A-normal form; (b) insertion of dynamic checks; and (c) insertion of
bounds widening expressions.
A-normal form conversion is straightforward: compound expressions are
handled by storing results of subexpressions into temporary variables,
as in the following example.

{\vspace*{-0.5em}
{\small
\begin{center}
$
\begin{array}{l}
$\code{let y=(x+1)+(6+1)}$
\;
\begin{frame}

\tikz\draw[-Latex,line width=2pt,color=orange] (0,0) -- (1,0);

\end{frame}
\;
\begin{array}{l}
$\code{let a=x+1;}$\\
$\code{let b=6+1;}$\\
$\code{let y=a+b}$\\
\end{array}
\end{array}
$
\end{center}
}
}

This simplifies the management of effects from subexpressions. The
next two steps of compilation are more interesting.

\begin{figure}[t!]
  \begin{small}
\begin{lstlisting}[mathescape,xleftmargin=4 mm]
/* p : $\color{purple!40!black}\tntarrayptr{0}{0}{\tint}{\cmode}$  */
/* $\color{purple!40!black}\rho$(p) = p_lo,p_hi */
{
  let x = strlen(p);
  if (x > 1) putchar(*(p+1));
}
\end{lstlisting}
\begin{frame}

\tikz\draw[-Latex,line width=2pt,color=orange] (0,0) -- (1,0);

\end{frame}
\begin{lstlisting}[xleftmargin=4 mm]
{
  assert(p_lo <= 0 && 0 <= p_hi); // bounds check
  assert(p != 0); // null check
  let x = strlen(p);
  let p_hi_new = x;
  p_hi = max(p_hi, p_hi_new);
  if (x > 1) {
    assert(p != 0); // null check for p + 1
    let p_1 = p + 1;
    assert(p_lo <= 1 && // bounds check for p + 1
     1 <= p_hi);    
    putchar(*p_1);
  }
}
\end{lstlisting}
\end{small}
\caption{Compilation Example for Check Insertions
}
\label{fig:compilationexample}
\end{figure}

\begin{figure}[t!]
  \begin{small}
\begin{lstlisting}[mathescape,xleftmargin=4 mm]
int deref_array(n : int,
     p :  $\color{green!40!black}\tntarrayptr{0}{n}{\tint}{\cmode}$) {
  /* $\color{purple!40!black}\rho$(p) = p_lo,p_hi */
  if (* p)
    (* (p + 1))
    else 0
}
...
/* p0 : $\color{purple!40!black}\tntarrayptr{0}{5}{\tint}{\cmode}$ */
deref_array(5, p0);
    \end{lstlisting}
\begin{frame}

\tikz\draw[-Latex,line width=2pt,color=orange] (0,0) -- (1,0);

\end{frame}
\begin{lstlisting}[xleftmargin=4 mm]
deref_array(n, p) {
  let p_lo = 0;
  let p_hi = n;
  /* runtime checks */
  assert(p_lo <= 0 && 0 <= p_hi);
  assert(p != 0);
  let p_derefed = *p;
  if (p_derefed != 0) {
    /* widening */
    if (p_hi == 0) {
      p_hi = p_hi + 1;
    }
    /* null check before pointer arithmetic */
    assert(p != 0);
    let p0 = p + 1;
    assert(p_lo <= 1 && 1 <= p_hi);
    (* p0)
  }
  else {
    0
  }
}
...
deref_array(5, p0);
    \end{lstlisting}
\end{small}
    \caption{Compilation Example for Dependent Functions}
\label{fig:compilationexample1}
\end{figure}


During compilation, $\Gamma$ tracks the lower and upper bound
associated with every pointer variable according to its type. At each
declaration of a \code{nt_array_ptr} variable \code{p}, the
compiler allocates two \emph{shadow variables}, 
stored in $\rho(p)$; these are initialized to \code{p}'s declared bounds
and will be updated during bounds widening.\footnote{Shadow variables
  are not used for \code{array_ptr} types (the bounds expressions are)
  since they are not subject to bounds widening.} 
Fig.~\ref{fig:compilationexample} shows how an invocation of
\code{strlen} on a null-terminated string is compiled into C
code. Each dereference of a checked pointer requires a null check
(See \textsc{S-DefNull} in Fig.~\ref{fig:semantics}), which the
compiler makes explicit: Line~$3$ of the generated code has the null
check on pointer \code{p} due to the \code{strlen},
  and a similar check happens
  at line~$8$ due to the pointer arithmetic on \code{p}.
Dereferences also require bounds checks: line~$2$ checks \code{p} is
in bounds before computing \code{strlen(p)}, while line~$10$ does
likewise before computing \code{*(p+1)}.

For \code{strlen(p)} and conditionals \code{if(*p)}, the \lang
semantics allows the upper bound of \code{p} to be extended.
The compiler explicitly inserts statements to do so on \code{p}'s
shadow bound variables. For example,
Fig.~\ref{fig:compilationexample}~line~$6$ widens \code{p}'s upper
bound if \code{strlen}'s result is larger than the existing
bound. 
Lines 7--12 of the generated code in
Fig.~\ref{fig:compilationexample1}
show how bounds are 
widened when compiling expression \code{if(*p)}. If we find that the
current \code{p}'s relative upper bound is equal to $0$ (line 10),
and \code{p}'s content is not null (line 8), we then increase the
upper bound by $1$ (line 11).

Fig.~\ref{fig:compilationexample1} also shows a dependent function call.
Notice that the bounds for the array pointer \code{p} are not passed as
arguments. Instead, they are initialized according to \code{p}'s
type---see line~3 of the original \lang program at the top of the figure.
Line~$2$ of the generated code
sets the lower bound  to \code{0} and line~$3$ sets the
upper bound to \code{n}.

\subsection{Comparison with Checked C Specification}
\label{sec:disc}

\begin{figure}[t]
{\small
{\captionsetup[lstlisting]{margin = 8 mm}
  \begin{lstlisting}[xleftmargin=8 mm]
nt_array_ptr<char> safe_strcat_c 
   (nt_array_ptr<char> dst : count(n),
    nt_array_ptr<char> src : count(0), int n) {
  nt_array_ptr<char> tmp : count(n) = dst;
  int x = strlen(tmp);
  /* tmp now has x as its upper bound */
  /* dst still has n as its upper bound */
  int y = strlen(src);

  if (x+y < n) {
    for (int i = 0; i < y; ++i)
      *(dst+x+i) = *(src+i);
    *(dst+x+y) = '\0';
    return dst;
  }
  return null;
}
\end{lstlisting}
}
}
\caption{Safe \code{strcat} in Checked C that avoids a run-time error
  exhibited by \code{safe_strcat} (Fig.~\ref{fig:strcat-ex}) when
  compiled with the current Checked C compiler}
\label{fig:strcatc-ex}
\end{figure}

The use of shadow variables for bounds widening is a key novelty of our
compilation approach, and adds more precision to bounds checking at runtime
compared to the official specification and current implementation of
\checkedc~\cite[5.1.2, pg 85]{checkedc}.  For example, the
\code{safe_strcat} example of Fig.~\ref{fig:strcat-ex} compiles with the
current Clang \checkedc compiler but will fail with a runtime
error. The statement \code{int x = strlen(dst)} at line 4 changes the
statically determined upper bound of \code{dst} to \code{x}, which can be smaller than
\code{n}, the full capacity of \code{dst}. The attempt to recover
the full capacity of \code{dst} through a dynamic cast
at line 7 will always fail if the capacity \code{n} is checked
against the statically determined new upper bound \code{x}. 
This problem can be worked around by invoking \code{strlen} on a temporary
variable \code{tmp} instead of \code{dst}
as in \code{safe_strcat_c} in Fig.~\ref{fig:strcatc-ex} (lines 4-5).
Likewise, if we were to add line \code{putchar(*(p+1));} 
after line 6 in the original code at the top of
Fig.~\ref{fig:compilationexample}, the code will always fail: the Clang \checkedc
compiler (with the transliterated C code as its input) would 
check \code{p} against its \emph{original} bounds \code{(0,0)} since the
updated upper bound \code{x} is now out of
the scope. Shadow variables address these problems because
they retain widened bounds beyond the scope of variables that store
them (i.e., \code{x} in both examples).

To make it match the specification, our compilation definition could
easily eschew
shadow variables and rely only on the type-based
bounds expressions available in $\Gamma$ for checking. However, doing so would 
force us to weaken the simulation theorem, reduce expressiveness,
and/or force the semantics to be more awkward. We plan to work with
the \checkedc team to implement our approach in a future revision.


\subsection{Metatheory}
\label{sec:meta}


We formalize both the compilation procedure and the simulation
theorem in the PLT Redex model we developed for \lang (see Sec.~\ref{sec:syntax}),
and then attempt to falsify it via Redex's support for random
testing. Redex allows us
  to specify compilation as logical rules (essentially, an extension
  of typing), but then execute it algorithmically to
  automatically test whether simulation holds. This process revealed
  several bugs in compilation and the theorem statement.
%
We ultimately plan to prove simulation in the Coq model.

We use the notation $\gg$ to
indicate the \emph{erasure} of stack and heap---the rhs is the same as
the lhs but with type annotations removed:
\begin{equation*}
  \begin{split}
    \heap  \gg & \dot \heap \\
    \varphi \gg & \dot \varphi
  \end{split}
\end{equation*}
In addition, when $\Gamma;\emptyset\vdash
\varphi$ and $\varphi$ is well-formed, we write $(\varphi,\heap,e) \gg (\dot \varphi, \dot \heap,
\dot e)$ to denote $\varphi \gg \dot \varphi$, $\heap \gg \dot \heap$
and $\Gamma;\Theta;\emptyset \vdash e \gg \dot e : \tau$ for some $\tau$ respectively. $\Gamma$ is omitted from the notation since the well-formedness of $\varphi$ and its consistency with respect to $\Gamma$ imply that $e$ must be closed under $\varphi$, allowing us to recover $\Gamma$ from $\varphi$.
Finally, we use $\xrightarrow{\cdot}^*$ to denote the transitive closure of the
reduction relation of $\elang$. Unlike the $\lang$, the semantics of
$\elang$ does not distinguish checked and unchecked regions.

Fig.~\ref{fig:checkedc-simulation-ref} gives an overview of 
the simulation theorem.\footnote{We ellide the  possibility of $\dot e_1$ evaluating to $\ebounds$ or $\enull$ in the diagram for readability.} The simulation theorem is specified in a way
that is similar to the one by~\citet{merigoux2021catala}.
An ordinary simulation property would
replace the middle and bottom parts of the figure with the
following: \[(\dot \varphi_0, \dot \heap_0, \dot e_0) 
  \xrightarrow{\cdot}^* (\dot \varphi_1, \dot \heap_1, \dot e_1)\]
Instead, we relate two erased configurations using the relation $\sim$,
which only requires that the two configurations will eventually reduce
to the same state. We formulate our simulation theorem differently
because the standard simulation theorem imposes a very strong
syntactic restriction to the compilation strategy. Very often, $(\dot
\varphi_0, \dot \heap_0, \dot e_0)$ reduces to a term that is
semantically equivalent to $(\dot \varphi_1, \dot \heap_1, \dot e_1)$,
but we are unable to syntactically equate the two configurations due
to the extra binders generated for dynamic checks and ANF
transformation. In earlier versions of the Redex model, we attempted
to change the compilation rules so the configurations could match
syntactically. However, the approach scaled poorly as we added
additional rules. 
This slight relaxation on the equivalence relation
between target configurations allows us to specify compilation more
naturally without having to worry about syntactic constraints.

\begin{thm}[Simulation ($\sim$)]\label{simulation-thm}
For \lang expressions $e_0$, stacks $\varphi_0$, $\varphi_1$, and heap snapshots $\heap_0$, $\heap_1$, 
if $\heap_0 \vdash \varphi_0$, $(\varphi_0,\heap_0,e_0)\gg(\dot \varphi_0,\dot \heap_0, \dot e_0)$,
and if there exists some $r_1$ such that $(\varphi_0, \heap_0, e_0)
\rightarrow_\cmode (\varphi_1, \heap_1, r_1)$, then the following facts hold:

\begin{itemize}

\item if there exists $e_1$ such that $r=e_1$ and $(\varphi_1, \heap_1, e_1) \gg (\dot \varphi_1, \dot \heap_1, \dot e_1)$, then there exists some $\dot \varphi$,$\dot \heap$, $\dot e$, such that
$(\dot \varphi_0, \dot \heap_0,\dot e_0) \xrightarrow{\cdot}^* (\dot
\varphi,\dot \heap,\dot e)$ and $(\dot
\varphi_1,\dot \heap_1,\dot e_1) \xrightarrow{\cdot}^* (\dot \varphi,
\dot \heap,\dot e)$.

\item if $r_1 = \ebounds$ or $\enull$, then we have $(\dot \varphi_0, \dot \heap_0,\dot e_0) \xrightarrow{\cdot}^* (\dot
\dot \varphi_1,\dot \heap_1, r_1)$ where $\varphi_1 \gg \dot
\varphi_1$, $\heap_1 \gg \dot \heap_1$.

\end{itemize}
\end{thm}

Our random generator (discussed in the next section) never produces
$\euncheckedtext$ expressions (whose behavior could be undefined), so we
can only test a the simulation theorem 
as it applies to checked code. This limitation makes it
unnecessary to state the other direction of the simulation theorem
where $e_0$ is stuck, because Theorem~\ref{thm:progress} guarantees
that $e_0$ will never enter a stuck state if it is well-typed in
checked mode.

The current version of the Redex model has been tested against $20000$
expressions with depth less than $10$. Each expression can
reduce multiple steps, and we test simulation between every two
adjacent steps to cover a wider range of programs, particularly the
ones that have a non-empty heap.

\begin{figure}[t]
{\small
\[
\begin{array}{c}
\begin{tikzpicture}[
            > = stealth, 
            shorten > = 1pt, 
            auto,
            node distance = 3cm
        ]

\begin{scope}[every node/.style={draw}]
    \node (A) at (0,1.5) {$\varphi_0,\heap_0, e_0$};
    \node (B) at (4,1.5) {$\varphi_1, \heap_1 ,e_1$};
    \node (C) at (0,0) {$\dot \varphi_0, \dot \heap_0 ,\dot e_0$};
    \node (D) at (4,0) {$\dot \varphi_1, \dot \heap_1, \dot e_1$};
    \node (E) at (2,-1.5) {$\dot \varphi,\dot \heap ,\dot e$};
\end{scope}
\begin{scope}[every edge/.style={draw=black}]

    \path [->] (A) edge node {$\longrightarrow_\cmode$} (B);
    \path [<->] (A) edge node {$\gg$} (C);
    \path [<->] (B) edge node {$\gg$} (D);
    \path [dashed,<->] (C) edge node {$\sim$} (D);
    \path [dashed,->] (C) edge node {$\xrightarrow{\cdot}^*$} (E);
    \path [dashed,->] (D) edge node[above] {$\xrightarrow{\cdot}^*$} (E);
\end{scope}

\end{tikzpicture}
\end{array}
\]
}
\caption{Simulation between \lang and \elang }
\label{fig:checkedc-simulation-ref}
\end{figure}

%% file: evaluation.tex
\section{Random Testing via the Implementation}\label{sec:evaluation}


In addition to using the \lang Redex model to establish simulation of
compilation (Section~\ref{sec:meta}), we also used it to gain confidence
that our model matches the Clang \checkedc implementation.
Disagreement on outcomes signals a bug in either the model or the
compiler itself. Doing so allowed us to quickly iterate on the
design of the model while adding new features, and revealed
several bugs in the Clang \checkedc implementation.

\myparagraph{Generating Well Typed Terms} For this random generation, we follow
the approach of~\citet{PalkaAST11} to generate well-typed \checkedc
terms by viewing the typing rules as generation rules.
%
Suppose we have a context $\Gamma$, a mode
$m$ and a type $\tau$, and we are
trying to generate a well-typed expression. We can do that by
reversing the process of type checking, selecting a typing rule and
building up an expression in a way that satisfies the rule's premises.

Recall the typing rule for dereferencing an
array pointer, which we depict below as
\textsc{G-DefArr}\footnote{Generator rules G-* correspond one to one
  with the type rules T-* in Sec.~\ref{sec:type-system}.}, color-coded to represent \textcolor{red}{inputs} and \textcolor{blue}{outputs} of the
generation process:%
\footnote{This input-output marking is commonly called a mode in
  the literature, but we eschew this term to avoid
  confusion with our pointer mode annotation.}
%
\begin{mathpar}
    \inferrule[G-DefArr]
              {\textcolor{red}{\Gamma};\textcolor{red}{\Theta} \vdash_{\textcolor{red}{m}} \textcolor{blue}{e} : \tptr{\tallarraybc{\textcolor{cyan}{\bvar}}{\textcolor{red}{\tau}}}{\textcolor{cyan}{m'}} \\
\textcolor{red}{m} \leq \textcolor{cyan}{m'} 
              }
              {\textcolor{red}{\Gamma};\textcolor{red}{\Theta} \vdash_{\textcolor{red}{m}} \textcolor{blue}{\estar{e}} : \textcolor{red}{\tau}}
%
  \end{mathpar}
%
If we selected \textsc{G-DefArr} for generating an expression, the
generated expression has to have the form $\textcolor{blue}{*e}$,
for some $\textcolor{blue}{e}$, to be generated according
to the rule's premises.
To satisfy the premise
$\textcolor{red}{\Gamma};\textcolor{red}{\Theta}
\vdash_{\textcolor{red}{m}} \textcolor{blue}{e} :
\tptr{\tallarraybc{\textcolor{cyan}{\bvar}}{\textcolor{red}{\tau}}}{\textcolor{cyan}{m'}}$,
we essentially need to make a recursive call to the generator, with
appropriately adjusted inputs.
However, the type in this judgment is not fixed yet---it contains
three unknown variables: $\textcolor{cyan}{m'}$,
$\textcolor{cyan}{\bvar}$, and $\textcolor{cyan}{\kappa}$---that need
to be generated before making the call.
Looking at the second premise informs that generation:
if the input mode $\textcolor{red}{m}$ is $\umode$, then $\textcolor{cyan}{m'}$
needs to be $\umode$ as well; if not, it is unconstrained,
just like $\textcolor{cyan}{\bvar}$ and $\textcolor{cyan}{\kappa}$, and therefore all three are free to be generated at random.
Thus, the recursive call to generate $\textcolor{blue}{e}$ can now
be made, and the \textsc{G-DefArr} rule returns $\textcolor{blue}{*e}$
as its output.


Using such generator rules, we can create a generator for random
well-typed terms of a given type in a straightforward manner: find all
rules whose conclusion matches the given type and then randomly choose
a candidate rule to perform the generation. To ensure that this
process terminates, we follow the standard practice of using ``fuel'' to
bound the depth of the generated terms; once the fuel is exhausted,
only rules without recursive premises are
selected~\cite{Pierce:SF4}. Similar methods were used for
generating top level functions and \kw{struct} definitions. 


While using just the typing-turned-generation rules is in theory
enough to generate all well-typed terms, it's more effective in
practice to try and exercise interesting patterns.  As
in~\citet{PalkaAST11} this can be viewed as a way of adding admissible
but redundant typing rules, with the sole purpose of
using them for generation. For example, below is one such rule,
\textsc{G-ASTR}, which creates an initialized null-terminated string that
is statically cast into an array with bounds $(0,0)$.
%
  \begin{mathpar}
    \inferrule[G-ASTR]
    {
      \textcolor{cyan}{i} \in \mathbb{N}^* \quad\quad
      \textcolor{cyan}{n_0,\ldots,n_{i-1} } \in \mathbb{Z} \quad\quad
      \texttt{fresh}(\textcolor{cyan}{x})\\
      \textcolor{red}{\Gamma} \vdash_{\textcolor{red}{m}}
      \textcolor{blue}{e'} :
      \tptr{\textcolor{red}{\tntarray{0}{\textcolor{cyan}{i}}{\tint}}}{c}
      \\      
      \textcolor{blue}{e} = \elet{\textcolor{cyan}{x}}{\textcolor{blue}{e'}}{\texttt{(init $\textcolor{cyan}{x}$ with }\textcolor{cyan}{n_0,\ldots
        n_{i-1}}\texttt{)};\textcolor{cyan}{x}}
}
    {\textcolor{red}{\Gamma} \vdash_{\textcolor{red}{m}} \textcolor{blue}{\ecast{\tptr{\tntarray{0}{0}{\tint}}{c}}{e}} : \textcolor{red}{\tptr{\tntarray{0}{0}{\tint}}{c}}}
  \end{mathpar}
%
Given some positive number $\textcolor{cyan}{i}$, numbers
$\textcolor{cyan}{n_0, \ldots, n_{i-1}}$, and a fresh variable
$\textcolor{cyan}{x}$ (which are arbitrarily generated), we can
recursively generate a pointer $\textcolor{blue}{e'}$ with bounds
$(0,\textcolor{cyan}{i})$, and initialize it with the generated $n_j$
using $\textcolor{cyan}{x}$ to temporarily store the pointer.

This rule is particularly useful when combined with \textsc{G-IfNT}
since there is a much higher chance of obtaining a non-zero value when
evaluating $\estar{p}$ in the guard of $\eiftext$, skewing the
distribution towards programs that enter the $\ethentext$ branch.
Relying solely on the type-based rules, entering the $\ethentext$
branch requires that \textsc{G-AssignArr} was chosen before \textsc{G-IfNT},
and that assignment would have to appear before $\eiftext$, which
means additional \textsc{G-Let} rules would need to be chosen: this
combination would therefore be essentially impossible to generate in
isolation.

Adding admissible generation rules like \textsc{G-Astr} in this
manner, as described in~\citet{PalkaAST11}, is a manual process. It is
guided by gathering statistics on the generated data and focusing on
language constructs that appear underrepresented in the posterior
distribution. For example, we arrived at the \textsc{G-Astr} rule by
recognizing that the pure type-based generation was not generating
non-trivial null-terminated strings, and then analyzing the sequence
of random choices that could lead to their generation.

\myparagraph{Generating Ill-typed Terms}
We can use generated well-typed terms to test our simulation theorem
(Section~\ref{sec:compilation}) and test that \lang and \checkedc
Clang agree on what is type-correct. But it is also useful to generate
ill-typed terms to test that \lang and \checkedc Clang also agree on
what is not.  However, while it is easy to generate arbitrary
ill-typed terms, they would be very unlikely to trigger any
inconsistencies; those are far more likely to exist on the boundary
between well- and ill-typedness. Therefore, we also manually added
variations of existing generation rules modified to be slightly more
permissive, e.g., by relaxing a single premise, thus allowing terms
that are ``a little'' ill-typed to be generated. Unlike coming up with
admissible generation rules like \textsc{G-Astr} (which is quite
challenging to automate), systematically and automatically relaxing
premises of existing rules seems feasible, and worthwhile future work.

\myparagraph{Random Testing for Language Design}
We used our Redex model and random generator to successfully guide the
design of our formal model, and indeed the Clang Checked C
implementation itself, which is being actively developed. To that end, we implemented a
conversion tool that converts \lang into a subset of the \checkedc
language and ensured that model and implementation exhibit the same
behavior (accept and reject the same programs and yield the same return
value).

This approach constitutes an interesting twist to traditional
model-based checking approaches.  Usually, one checks that the
implementation and model agree on all inputs \emph{of the
  implementation}, with the goal of covering as many behaviors as
possible. This is the case, for example, in \citet{lambdajs}, where
they use real test suites to demonstrate the faithfullness of their
core calculus to Javascript. Our approach and goal in this work is
essentially the opposite: as the Clang Checked C implementation does
not fully implement the Checked C spec, there is little hope of
covering all terms that are generated by Clang Checked C. Instead,
we're looking for \emph{inconsistencies}, which could be caused by
bugs either in the Clang Checked C compiler or our own model.  

One inconsistency we found comes from the following:

{\small 
\begin{lstlisting}[xleftmargin=4 mm]
 array_ptr<char> fun(void) : count(3) {
   array_ptr<char> x : count(3);
   x = calloc(3, sizeof(char));
   return x+3;
 }
 int main(void) {
   *(fun()) = 0;
   return 0;
 }
\end{lstlisting}
}%
\noindent
In this code, the function \code{fun} is supposed to return a
checked array pointer of size 3. Internally, it allocates such an
array, but instead of returning the pointer \code{x} to that array, it
increments that pointer by 3. Then, the \code{main} function just
calls \code{fun}, and tries to assign 0 to its result. Our model
correctly rules out this program, while the Clang Checked C
implementation happily accepted this out-of-bounds
assignment. Interestingly, it correctly rejected programs where the
array had size 1 or 2. This inconsistency has been fixed in the latest
version of the compiler.

We also found the opposite kind of inconsistency---programs that
the Clang Checked C implementation rejects contrary to the spec.
For instance:%
\footnote{After minimization, this turned out to be a known issue: 
  \url{https://github.com/microsoft/checkedc-clang/issues/1008}}

{\small 
\begin{lstlisting}[xleftmargin=4 mm]
array_ptr<int> f(void) : count(5) {
  array_ptr<int> x : count(5) =
    calloc<int>(5, sizeof(int)); 
  return x;
}
array_ptr<int> g(void ) : count(5) {
  array_ptr<int> x : count(5) =
    calloc<int>(5, sizeof(int)); 
  return x+3;
}
int main(void) {
  return *(0 ? g() : f() + 3);
}    
\end{lstlisting}
}%
\noindent
In this piece of code both \code{f} and \code{g} functions compute a
pointer to the same index in an array of size 5 (as \code{f} calls
\code{g}). The \code{main} function then creates a ternary expression
whose branches call \code{f} and \code{g}, but the Clang Checked C
implementation rejects this program, as its static analysis is not
sophisticated enough to detect that both branches have the same type.

%% file: comparison.tex
\section{Related Work}
\label{sec:related}

Our work is most closely related to prior formalizations of C(-like)
languages that aim to enforce memory safety, but it also touches on
C-language formalization in general.

\myparagraph{Formalizing C and Low-level code}
A number of prior works have looked at formalizing the semantics of C,
including CompCert~\cite{Blazy2009,leroy:hal-00703441},
\citet{ellison-rosu-2012-popl}, \citet{Kang:2015:FCM:2813885.2738005},
and \citet{10.1145/2980983.2908081, Memarian:2019:ECS:3302515.3290380}. These works also model
pointers as logically coupled with either the bounds of the blocks
they point to, or provenance information from which bounds can be
derived. None of these is directly concerned with enforcing
spatial safety, and that is reflected in the design. For example,
memory itself is not be represented as a flat address space, as in our
model or real machines, so memory corruption due to spatial safety
violations, which Checked C's type system aims to prevent, may not be
expressible. That said, these formalizations consider much more of the
C language than does \lang, since they are interested in the entire
language's behavior.

\myparagraph{Spatially Safe C Formalizations}
Several prior works formalize C-language transformations or C-language
dialects aiming to ensure spatial safety. 
\citet{10.1145/2813885.2737979} extends the formalization
of \citet{ellison-rosu-2012-popl} to produce a semantics that detects
violations of spatial safety (and other forms of undefinedness). It
uses a CompCert-style memory model, but ``fattens'' logical pointer
representations to facilitate adding side conditions similar to \lang's.
Its concern is bug finding, not compiling programs to
use this semantics.

CCured~\cite{Necula2005} and Softbound~\cite{softbound} implement
spatially safe semantics for normal C via program transformation. Like
\lang, both systems' operational semantics annotate pointers with
their bounds. CCured's equivalent of array pointers are compiled to be
``fat,'' while SoftBound compiles bounds metadata to a separate
hashtable, thus retaining binary compatibility at higher checking
cost. Checked C uses static type information to enable bounds checks
without need of pointer-attached metadata, as we show in
Section~\ref{sec:compilation}. Neither CCured nor Softbound models
null-terminated array pointers, whereas our semantics ensures that
such pointers respect the zero-termination invariant, leveraging
bounds widening to enhance expressiveness.

Cyclone \cite{Jim2002,GrossmanMJHWC02} is a C dialect that aims to
ensure memory safety; its pointer types are similar to
CCured. Cyclone's formalization~\cite{GrossmanMJHWC02} focuses on the
use of \emph{regions} to ensure temporal safety; it does not formalize
arrays or threats to spatial safety. Deputy~\cite{Feng2006,Condit2007}
is another safe-C dialect that aims to avoid fat pointers; it was an
initial inspiration for Checked C's design~\cite{Elliott2018}, though
it provides no specific modeling for null-terminated array
pointers. Deputy's formalization~\cite{Condit2007} defines its
semantics directly in terms of compilation, similar in style to what
we present in Section~\ref{sec:compilation}. Doing so tightly couples
typing, compilation, and semantics, which are treated independently in
\lang. Separating semantics from compilation isolates meaning 
from mechanism, easing understandability. Indeed, it was this
separation that led us to notice the 
limitation with Checked C's handling of bounds widening.

The most closely related work is the
formalization of \checkedc done by \citet{ruef18checkedc-incr}. They
present the type system and semantics of a core model of \checkedc,
mechanized in Coq, and were the first to prove a blame theorem.  \lang's
Coq-based development (Section~\ref{sec:formal}) substantially extends
theirs to include conditionals, dynamically bounded array
pointers with dependent types, null-terminated array pointers, 
dependently typed functions, and subtyping. They postulate that pointer metadata can
be erased in a real implementation, but do not show it. Our \CoreChkC
compiler, formalized and
validated in PLT Redex via randomized testing, demonstrates that such
metadata \emph{can} be erased; we found that erasure was non-obvious once
null-terminated pointers and bounds widening were considered.

%% file: conclusion.tex
\section{Conclusion and Future Work}
\label{sec:conclude}

This paper presented \CoreChkC, a formalization of an extended core of
the Checked C language which aims to provide spatial memory
safety. \CoreChkC models dynamically sized and
null-terminated arrays with dependently typed bounds that can
additionally be widened at runtime. We prove, in Coq, the key safety
property of Checked C for our formalization, {\em blame}: if a mix of
checked and unchecked code gives rise to a spatial memory safety
violation, then this violation originated in an unchecked part of the
code. We also show how programs written in \CoreChkC (whose
semantics leverage fat pointers) can be compiled to \elang (which does
not) while preserving their behavior. We developed a version of \lang
written in PLT Redex, and used a custom term generator in conjunction with Redex's
randomized testing framework to give confidence that compilation is
correct. We also used this framework to cross-check \lang
against the \checkedc compiler, finding multiple inconsistencies
in the process. 

As future work, we wish to extend \CoreChkC to model more of Checked
C, with our Redex-based testing framework guiding the process. The
most interesting Checked C feature not yet modeled is \emph{interop
  types} (itypes), which are used to simplify interactions with
unchecked code via function calls. A function whose
parameters are itypes can be passed checked or unchecked pointers
depending on whether the caller is in a checked region. This feature
allows for a more modular C-to-Checked C porting process, but complicates reasoning
about blame. A more ambitious next step would be to extend an existing
formally verified framework for C, such as CompCert~\cite{compcert} or VeLLVM
\cite{Zhao:2012:FLI:2103621.2103709}, with Checked C features, towards
producing a verified-correct Checked C compiler. We believe that
\lang's Coq and Redex models lay the foundation for such a step, but
substantial engineering work remains.

%% file: appendix.tex
\section{Appendix}

\subsection{Differences with the Coq and Redex Models}\label{app:model-diffs}


The Coq and Redex models of \lang may be found at \url{https://github.com/plum-umd/checkedc}.
The Coq model's syntax is slightly different from that in
Fig.~\ref{fig:checkc-syn}. In particular, the arguments in a function call
are restricted to variables and constants, according to
a separate well-formedness condition. A function call \code{f(e)} can
always be written in \code{let x = e in f(x)} to cope.
In addition,
conditionals have two syntactic forms: \code{EIf} is a normal
conditional, and \code{EIfDef} is one whose boolean guard is of the
form \estar{x}. By syntactically distinguishing these two
cases, the Coq model does not need the \emph{[prefer]} rule for
\code{if (*x)}$...$ forms as in
Fig.~\ref{fig:c-context}. The Redex model \emph{does} prioritize such
forms but not the same way as in the figure. It uses a variation of
the \textsc{S-Var} rule: The 
modified rule is equipped with a precondition that is false whenever
\textsc{S-IfNTT} is applicable.

The Coq model uses a runtime stack $\varphi$ as described at the start
of Sec.~\ref{sec:semantics}.
The Redex model introduces let bindings during evaluation to
simulate a runtime stack. For example, consider the expression 
$e \equiv \elet{x}{(\evalue{5}{\tint})}{\ebinop{x}{x}}$. Expression $e$ first steps to
$\elet{x}{(\evalue{5}{\tint})}{\ebinop{(\evalue{5}{\tint})}{x}}$, which in
turns steps to
$\elet{x}{(\evalue{5}{\tint})}{\ebinop{(\evalue{5}{\tint})}{(\evalue{5}{\tint})}}$. Since
the rhs of $x$ is a value, the let binding in $e$ effectively
functions as a stack that maps from $x$ to $\evalue{5}{\tint}$. The
let form remains in the expression and lazily replaces the variables
in its body. The let form can be removed from the expression only if its
body is evaluated to a value, e.g., 
$\elet{x}{(\evalue{5}{\tint})}{(\evalue{10}{\tint})}$
steps to 
$\evalue{10}{\tint}$. The rule for popping let bindings in this manner
corresponds to the \textsc{S-Ret} rule in Fig.~\ref{fig:semantics}.
Leveraging let bindings adds complexity to the semantics but
simplifies typing/consistency and term generation during randomized testing.

\subsection{Typing Rules for Literal Pointers}\label{sec:literal-pointer-typing}

The typing of integer literals, which can also be pointers to the
heap, was presented in Sec.~\ref{sec:theorem} in Fig.~\ref{fig:const-type}. Here
we describe these rules further.

The variable type rule
($\textsc{T-Var}$) simply checks if a given variable has the defined
type in $\Gamma$; the constant rule ($\textsc{T-Const}$) is slightly
more involved.  First, it ensures that the type annotation $\tau$ does
not contain any free variables. More importantly, it ensures that the
literal itself is well typed using an auxilliary typing relation
$\heap;\sigma \vdash n : \tau$.

If the literal's type is an integer, an unchecked pointer, or a null
pointer, it is well typed, as shown by the top three rules in
Fig.~\ref{fig:const-type}. However, if it is a checked pointer
$\tptr{\omega}{\cmode}$, we need to ensure that what it points to in
the heap is of the appropriate pointed-to type ($\omega$), and also
recursively ensure that any literal pointers reachable this way are
also well-typed. This is captured by the bottom rule in the figure,
which states that for every location $n+i$ in the pointers' range
$[n, n+\size(\omega))$, where $\size$ yields the size of its argument,
  then the value at the location $\heap(n+i)$ is also well-typed.
  However, as heap snapshots can contain cyclic structures (which
  would lead to infinite typing deriviations), we use a scope $\sigma$
  to assume that the original pointer is well-typed when checking the
  types of what it points to. The middle rule then accesses the scope
  to tie the knot and keep the derivation finite, just like in
  \citet{ruef18checkedc-incr}.

\subsection{Other Semantic Rules}\label{sec:rem-semantics}

Fig.~\ref{fig:rem-semantics} shows the remaining semantic rules for
$\lang$. We explain a selected few rules in this subsection.

\begin{figure*}[t]
{\small
\begin{mathpar}

\inferrule [S-Var]{} {(\varphi,\heap,x)\longrightarrow (\varphi,\heap,\varphi(x))}

    \inferrule[S-DefArray]{\heap(n)=\evalue{n_a}{\tau_a} \\ 0 \in [n_l,n_h)}
    {(\varphi,\heap,\estar{\evalue{n}{\tntarrayptr{n_l}{n_h}{\tau}{\cmode}}}) \longrightarrow (\varphi,\heap,\evalue{n_a}{\tau})}

    \inferrule[S-DefArrayBound]{0 \not\in [n_l,n_h)}
     { (\varphi,\heap,\estar{\evalue{n}{\tallarrayptr{n_l}{n_h}{\tau}{c}}}) \longrightarrow (\varphi,\heap,\ebounds)}

    \inferrule[S-DefNTArrayBound]{0 \notin [n_l,n_h]}
    {(\varphi,\heap,\estar{\evalue{n}{\tntarrayptr{n_l}{n_h}{\tau}{\cmode}}}) \longrightarrow (\varphi,\heap,\ebounds)}

    \inferrule[S-Assign]{\heap(n)=\evalue{n_a}{\tau_a} }
      {(\varphi,\heap,\eassign{\evalue{n}{\tptr{\tau}{\cmode}}}{\evalue{n_1}{\tau_1}}) \longrightarrow (\varphi,\heap[n \mapsto \evalue{n_1}{\tau}],\evalue{n_1}{\tau})}

    \inferrule[S-AssignNull]{}
      {(\varphi,\heap,\eassign{\evalue{0}{\tptr{\omega}{\cmode}}}{\evalue{n_1}{\tau_1}}) \longrightarrow (\varphi,\heap,\enull)}

    \inferrule[S-AssignArrBound]{0 \not\in [n_l,n_h) }
      {(\varphi,\heap,\eassign{\evalue{n}{\tallarrayptr{n_l}{n_h}{\tau}{\cmode}}}{\evalue{n_1}{\tau_1}}) \longrightarrow (\varphi,\heap,\ebounds)}

  \inferrule[S-Malloc]{\varphi(\omega)=\omega_a \\ \mathtt{alloc}(\heap,\omega_a)=(n,\heap')}
   { (\varphi,\heap,\emalloc{\omega}) \longrightarrow (\varphi,\heap',\evalue{n}{\tptr{\omega_a}{\cmode}})}

  \inferrule[S-MallocBound]{\varphi(\omega)=\tallarray{n_l}{n_h}{\tau}\\ (n_l \neq 0 \vee n_h \le 0)}
    { (\varphi,\heap,\emalloc{\omega}) \longrightarrow (\varphi,\heap',\ebounds)}

    \inferrule[S-IfT]{n \neq 0 }
    {(\varphi,\heap,\eif{\evalue{n}{\tau}}{e_1}{e_2}) \longrightarrow (\varphi,\heap,e_1)}

    \inferrule[S-IfF]{}
    {(\varphi,\heap,\eif{\evalue{0}{\tau}}{e_1}{e_2}) \longrightarrow (\varphi,\heap,e_2)}

    \inferrule[S-Unchecked]{}
    {(\varphi,\heap,\eunchecked{\evalue{n}{\tau}} \longrightarrow (\varphi,\heap,\evalue{n}{\tau})}

    \inferrule[S-Str]{
        0 \in [n_l,n_h]
\\ n_a \le n_h
\\ \heap(n+n_a) = 0 
\\ (\forall i. n \le i < n+n_a \Rightarrow (\exists n_i\;t_i. \heap(n+i) = \evalue{n_i}{\tau_i} \wedge n_i \neq 0))}
    {(\varphi,\heap,\estrlen{\evalue{n}{\tarrayptr{n_l}{n_h}{\tau}{m}}}) \longrightarrow (\varphi,\heap,\evalue{n_a}{\tint})}

    \inferrule[S-StrBounds]{
        0 \notin [n_l,n_h]
}
    {(\varphi,\heap,\estrlen{\evalue{n}{\tarrayptr{n_l}{n_h}{\tau}{c}}}) \longrightarrow (\varphi,\heap,\ebounds)}

    \inferrule[S-StrNull]{}
    {(\varphi,\heap,\estrlen{\evalue{0}{\tarrayptr{n_l}{n_h}{\tau}{c}}}) \longrightarrow (\varphi,\heap,\enull)}

    \inferrule[S-Add]{n = n_1 + n_2}
    {(\varphi,\heap,\evalue{n_1}{\tint} \plus \evalue{n_2}{\tint}) \longrightarrow (\varphi,\heap, n)}

    \inferrule[S-AddArr]{n = n_1 + n_2\\ n_l' = n_l - n_2 \\ n_h' = n_h - n_2}
    {(\varphi,\heap,\evalue{n_1}{\tallarrayptr{n_l}{n_h}{\tau}{m}} \plus \evalue{n_2}{\tint}) \longrightarrow (\varphi,\heap, \evalue{n}{\tallarrayptr{n_l'}{n_h'}{\tau}{m}})}

n    \inferrule[S-AddArrNull]{}
    {(\varphi,\heap,\evalue{0}{\tallarrayptr{n_l}{n_h}{\tau}{c}} \plus \evalue{n_2}{\tint}) \longrightarrow (\varphi,\heap, \enull)}

\end{mathpar}

}
\caption{Remaining \lang Semantics Rules (extends Fig.~\ref{fig:semantics})}
\label{fig:rem-semantics}
\end{figure*}

Rule \textsc{S-Var} loads the value for $x$ in stack $\varphi$.
Rule \textsc{S-DefArray} dereferences an array pointer, which is similar to the Rule \textsc{S-DefNTArray} in Fig.~\ref{fig:semantics} (dealing with null-terminated array pointers).
The only difference is that the range of $0$ is at $[n_l,n_h)$ not $[n_l,n_h]$, meaning that one cannot dereference the upper-bound position in an array.
Rules \textsc{DefArrayBound} and \textsc{DefNTArrayBound} describe an error case for a dereference operation.
If we are dereferencing an array/NT-array pointer and the mode is $\cmode$, $0$ must be in the range from $n_l$ to $n_h$ (meaning that the dereference is in-bound); if not, the system results in a $\ebounds$ error. Obviously, the dereference of an array/NT-array pointer also experiences a $\enull$ state transition if $n\le 0$.

Rules \textsc{S-Malloc} and \textsc{S-MallocBound} describe the $\emalloctext$ semantics. Given a valid type $\omega_a$ that contains no free variables, $\mathtt{alloc}$ function returns an address pointing at the first position of an allocated space whose size is equal to the size of $\omega_a$, and a new heap snapshot $\heap'$ that marks the allocated space for the new allocation. The $\emalloctext$ is transitioned to the address $n$ with the type ${\tptr{\omega_a}{\cmode}}$ and new updated heap. It is possible for $\emalloctext$ to transition to a $\ebounds$ error if the $\omega_a$ is an array/NT-array type $\tallarray{n_l}{n_h}{\tau}$, and either $n_l \neq 0$ or $n_h \le 0$. This can happen when the bound variable is evaluated to a bound constant that is not desired.

\subsection{Subtyping for dependent types}
\label{app:le}
  
The subtyping relation given in Fig.~\ref{fig:checkc-subtype} involves
dependent bounds, i.e., bounds that may refer to variables. To decide
premises $b \leq b'$, we need a decision procedure that accounts for
the possible values of these variables. This process considers
$\Theta$, tracked by the typing judgment, and $\varphi$, the current
stack snapshot (when performing subtyping as part of the type
preservation proof).

\begin{defi}[Inequality]

\begin{itemize}

\item $n \le m$ if $n$ is less than or equal to $m$.
\item $x+n \le x + m$ if $n$ is less than or equal to $m$.
\item All other cases result in $\efalse$.

\end{itemize}
\end{defi}

To capture bound variables in dependent types, the \checkedc subtyping
relation ($\sqsubseteq$) is parameterized by a restricted stack
snapshot $\varphi|_{\rho}$ and the predicate map $\Theta$, where
$\varphi$ is a stack and $\rho$ is a set of
variables. $\varphi|_{\rho}$ means to restrict the domain of $\varphi$
to the variable set $\rho$. Clearly, we have the relation:
$\varphi|_{\rho} \subseteq \varphi$. $\sqsubseteq$
being parameterized by $\varphi|_{\rho}$ refers to that when we
compare two bounds $b \le b'$, we actually do
$\varphi|_{\rho}(b) \le \varphi|_{\rho}(b')$ by interpreting the
variables in $b$ and $b'$ with possible values in $\varphi|_{\rho}$.
Let's define a subset relation $\preceq$ for two restricted stack
snapshot $\varphi|_{\rho}$ and $\varphi'|_{\rho}$:

\begin{defi}[Subset of Stack Snapshots]
  Given two $\varphi|_{\rho}$ and $\varphi'|_{\rho}$,
  $\varphi|_{\rho} \preceq \varphi'|_{\rho}$, iff for $x\in\rho$ and
  $y$,
  $(x,y) \in \varphi|_{\rho} \Rightarrow (x,y) \in \varphi'|_{\rho}$.
\end{defi}

For every two restricted stack snapshots $\varphi|_{\rho}$ and
$\varphi'|_{\rho}$, such that
$\varphi|_{\rho} \preceq \varphi'|_{\rho}$, we have the following
theorem in \checkedc (proved in Coq):

\begin{thm}[Stack Snapshot Theorem]
  Given two types $\tau$ and $\tau'$, two restricted stack snapshots
  $\varphi|_{\rho}$ and $\varphi'|_{\rho}$, if
  $\varphi|_{\rho}\preceq \varphi'|_{\rho}$, and
  $\tau \sqsubseteq \tau'$ under the parameterization of
  $\varphi|_{\rho}$, then $\tau \sqsubseteq \tau'$ under the
  parameterization of $\varphi'|_{\rho}$.
\end{thm}

Clearly, for every $\varphi|_{\rho}$, we have
$\emptyset \preceq \varphi|_{\rho}$. The type checking stage is a
compile-time process, so $\varphi|_{\rho}$
is $\emptyset$ at the type checking stage. Stack snapshots are needed
for proving type preserving, as variables in bounds expressions are
evaluated away.

\begin{figure}[h!]
{\small
  \begin{mathpar}
    \inferrule[T-Def]
              {\Gamma;\Theta \vdash_m e : \tptr{\tau}{m'} \\
                m \leq m'}
              {\Gamma;\Theta \vdash_m \estar{e} : \tau}

    \inferrule[T-Mac]
              {}
              {\Gamma; \Theta \vdash_m \emalloc{\omega} : \tptr{\omega}{\cmode}}

    \inferrule[T-Add]
              {\Gamma; \Theta \vdash_m e_1 : \tint \\
                \Gamma; \Theta \vdash_m e_2 : \tint}
              {\Gamma; \Theta \vdash_m (e_1 \plus e_2) : \tint }

    \inferrule[T-Ind] 
              {\Gamma; \Theta \vdash_m e_1 : \tptr{\tallarrayb{\bvar}{\tau}}{m'} \\
                \Gamma; \Theta \vdash_m e_2 : \tint \\
                m \leq m'}              
              {\Gamma; \Theta \vdash_m \estar{(\ebinop{e_1}{e_2})} : \tau}

    \inferrule[T-Assign]
              {\Gamma; \Theta \vdash_m e_1 : \tptr{\tau}{m'} \\
                \Gamma; \Theta \vdash_m e_2 : \tau' \\
                \tau'\sqsubseteq \tau \\
                m \leq m'}
              {\Gamma; \Theta \vdash_m \eassign{e_1}{e_2} : \tau}

   \inferrule[T-IndAssign]
              {\Gamma; \Theta \vdash_m e_1 : \tptr{\tallarrayb{\bvar}{\tau}}{m'}\\
                \Gamma; \Theta \vdash_m e_2 : \tint \\
                \Gamma; \Theta \vdash_m e_3 : \tau' \\
                \tau'\sqsubseteq \tau \\
                m \leq m'}
              {\Gamma; \defscope \vdash_m \eassign{(e_1 \plus e_2)}{e_3} : \tau}

  \end{mathpar}
}
\caption{Remaining \lang Type Rules (extends Fig.~\ref{fig:type-system-1})}
\label{fig:rem-type-system}
\end{figure}

As mentioned in the main text, $\sqsubseteq$ is also parameterized by
$\Theta$, which provides the range of allowed values for a bound
variable; thus, more $\sqsubseteq$ relation is provable. For example,
in Fig.~\ref{fig:strcat-ex}, the \code{strlen} operation in line 4
turns the type of \code{dst} to be $\tntarrayptr{0}{x}{\tint}{\cmode}$
and extends the upper bound to \code{x}. In the \code{strlen} type
rule, it also inserts a predicate \code{x}$\ge 0$ in $\Theta$; thus,
the cast operation in line 16 is valid because
$\tntarrayptr{0}{x}{\tint}{\cmode} \sqsubseteq
\tntarrayptr{0}{0}{\tint}{\cmode}$ is provable when we know
\code{x}$\ge 0$.

Note that if $\varphi$ and $\Theta$ are $\emptyset$, we do only the
syntactic $\le$ comparison; otherwise, we apply $\varphi$ to both
sides of $\sqsubseteq$, and then determine the $\le$ comparasion based
on a Boolean predicate decision procedure on top of $\Theta$. This
process allows us to type check both an input expression and the
intermediate expression after evaluating an expression. 

\subsection{Other Type Rules}\label{rem-type}

Here we show the type rules for other \checkedc operations in Fig.~\ref{fig:rem-type-system}.
Rule \textsc{T-Def} is for dereferencing a non-array pointer. 
The statement $m \leq m'$ ensures that no unchecked pointers are used in checked regions.
Rule \textsc{T-Mac} deals with
$\emalloctext$ operations. There is a well-formedness check to require
that the possible bound variables in $\omega$ must be in the domain of
$\Gamma$ (see Fig.~\ref{fig:wftypesandbounds}). This is similar to the well-formedness assumption of the type environment (Definition~\ref{type-wellformed}) Rule \textsc{T-Add} deals with binary operations whose sub-terms are integer expressions, while rule \textsc{T-Ind} serves the case for pointer arithmetic. For simplicity, in the \checkedc formalization, we do not allow arbitrary pointer arithmetic. The only pointer arithmetic operations allowed are the forms shown in rules \textsc{T-Ind} and \textsc{T-IndAssign} in Fig.~\ref{fig:rem-type-system}. Rule \textsc{T-Assign} assigns a value to a non-array pointer location. The predicate $\tau'\sqsubseteq \tau$ requires that the value being assigned is a subtype of the pointer type.
The \textsc{T-IndAssign} rule is an extended assignment operation for handling assignments for array/NT-array pointers with pointer arithmetic. Rule \textsc{T-Unchecked} type checks \code{unchecked} blocks.

\subsection{Struct Pointers}\label{appx:struct}

\checkedc has \kw{struct} types and \kw{struct} pointers. Fig.~\ref{fig:checkc-struct} contains the syntax of \kw{struct} types as well as new subtyping relations built on the \kw{struct} values.
For a \kw{struct} typed value, \checkedc has a special operation for it, which is $\eamper{e}{f}$. This operation indexes the $f$-th position \kw{struct}~$T$ item, if the expression $e$ is evaluated to a \kw{struct} pointer $\tptr{\tstruct{T}}{m}$. Rule \textsc{T-Struct} in Fig.~\ref{fig:checkc-struct} describes its typing behavior.
Rules \textsc{S-StructChecked} and \textsc{S-StructUnChecked} describe the semantic behaviors of $\eamper{e}{f}$ on a given \kw{struct} \code{checked}/\code{unchecked} pointers, while rule \textsc{S-StructNull} describes a \code{checked} \kw{struct} null-pointer case.
In our Coq/Redex formalization, we include the \kw{struct} values and the operation $\eamper{e}{f}$. We omit it in the main text due to the paper length limitation.

\begin{figure}
{\small
$\begin{array}{l}
\text{  Struct Syntax: }\\[0.5em]
  \begin{array}{ll}
 \mathtt{Type} & \tstruct{T}
\\[0.2em]
     \text{Structdefs} & D \; \in \; T \rightharpoonup fs \\[0.2em]
      \text{Fields} & fs \; ::= \; \tau~\mathtt{f} \mid \tau~\mathtt{f}; fs 
    \end{array}\\[2em]
\text{  Struct Subtype: }\\[0.5em]
\begin{array}{l}
    D(T) = fs \wedge fs(0) = \tnat \Rightarrow  \tptr{\tstruct{T}}{m} \sqsubseteq \tptr{\tnat}{m}\\[0.5em]
    D(T) = fs \wedge fs(0) = \tnat \wedge 0 \le b_l \wedge b_h \le 1 \\[0.2em]
 \qquad\qquad \Rightarrow 
       \tptr{\tstruct{T}}{m} \sqsubseteq \tarrayptr{b_l}{b_h}{\tnat}{m}
    \end{array}
\\[3em]
\text{  Struct Type Rule: }\\[0.5em]
    \end{array}
$

{
\begin{mathpar}
  \inferrule [T-Struct]
  {\Gamma; \Theta \vdash_m e :  {\tptr{\tstruct{T}}{m}} \\
    D(T) = fs \\ fs(f)=\tau_f}
  {\Gamma; \Theta \vdash_m \eamper{e}{f} : \tptr{\tau_f}{m}}

\end{mathpar}
}
{
$\begin{array}{l}
\text{Struct Semantics: }
\end{array}
$
}
{
\begin{mathpar}
  \inferrule [S-StructChecked]
  {n > 0 \\ D(T) = fs \\ fs(f)=\tau_a \\ n_a=\mathtt{index}(fs,f)}
  {(\varphi,\heap,\eamper{\evalue{n}{\tptr{\tstruct{T}}{\cmode}}}{f}) \longrightarrow (\varphi,\heap,\evalue{n_a}{\tptr{\tau_a}{\cmode}})}

  \inferrule [S-StructNull]
  {n = 0}
  {(\varphi,\heap,\eamper{\evalue{n}{\tptr{\tstruct{T}}{\cmode}}}{f}) \longrightarrow (\varphi,\heap,\enull)}

  \inferrule [S-StructUnChecked]
  {D(T) = fs \\ fs(f)=\tau_a \\ n_a=\mathtt{index}(fs,f)}
  {(\varphi,\heap,\eamper{\evalue{n}{\tptr{\tstruct{T}}{\umode}}}{f}) \longrightarrow (\varphi,\heap,\evalue{n_a}{\tptr{\tau_a}{\umode}})}

\end{mathpar}
}
}
  \caption{\lang Struct Definitions}
  \label{fig:checkc-struct}
\end{figure}

\begin{figure}[t]
{\small
  \begin{mathpar}

    \inferrule[]
    {}
    {\Gamma \vdash n}

    \inferrule[]
    {x:\tint \in \Gamma}
    {\Gamma \vdash x + n}

    \inferrule[]
    {\Gamma \vdash b_l\\
    \Gamma \vdash b_h}
  {\Gamma \vdash (b_l,b_h)}

  \inferrule[]
  {}
  {\Gamma \vdash \tint}

  \inferrule[]
  {\Gamma \vdash \bvar \\
  \Gamma \vdash \tau}
  {\Gamma \vdash \tptr{\tallarrayb{\bvar}{\tau}}{m}}

  \inferrule[]
  {\Gamma \vdash \tau}
  {\Gamma \vdash \tptr{\tau}{m}}

  \inferrule[]
  {T \in D}
  {\Gamma \vdash \tptr{\tstruct{T}}{m}}
  \end{mathpar}
}
 \caption{Well-formedness for Types and Bounds}
\label{fig:wftypesandbounds}
\end{figure}


\begin{figure}[t]
{\small
  \begin{mathpar}
    \inferrule[]
    {\Gamma \vdash \overline{x}:\overline{\tau} \\
      \Gamma[\overline{x} \mapsto \overline{\tau}] \vdash \tau \\
    \Gamma[\overline{x} \mapsto \overline{\tau}]; \Theta  \vdash_{\cmode} e : \tau}
    {\Gamma \vdash \tau\;(\overline{x}:\overline{\tau})\;e}

    \inferrule[]
    {}
    {\Gamma \vdash \cdot}

    \inferrule[]
    {\Gamma \vdash \tau \\
    \Gamma[x \mapsto \tau] \vdash \overline{x}:\overline{\tau}}
    {\Gamma \vdash x:\tau, \overline{x}:\overline{\tau}}
  \end{mathpar}
}
 \caption{Well-formedness for functions}
\label{fig:wffunctions}
\end{figure}

\begin{figure}[t]
{\small
  \begin{mathpar}
    \inferrule[]
    {\Gamma \vdash \tau}
    {\Gamma \vdash \tau~\mathtt{f} }

    \inferrule[]
    {\Gamma \vdash \tau\\
     \Gamma \vdash fs}
    {\Gamma \vdash \tau~\mathtt{f}; fs }

  \end{mathpar}
}
 \caption{Well-formedness for structs}
\label{fig:wfstructs}

{\small
  \begin{mathpar}

    \inferrule[]
    {\Gamma[\overline{x} \mapsto \overline{\tau}]; \emptyset  \vdash
      e \gg \dot e : \tau}
    {\Gamma \vdash \tau\;(\overline{x}:\overline{\tau})\;e \gg
      (\overline{x})\;\dot e}
  \end{mathpar}
}
 \caption{Compilation Rules for Functions}
\label{fig:compilefunctions}
\end{figure}

\subsection{The Compilation Rules}\label{appx:comp1}
Fig.~\ref{fig:syntaxerased} and Fig.~\ref{fig:semanticserased} shows
the syntax for $\elang$, the target language for compilation. We syntactically
restrict the expressions to be in A-normal form to simplify the
presentation of the compilation rules. In the Redex model, we
occasionally break this constraint to speed up the performance of
random testing by removing unnecessary let bindings.
To allow explicit runtime
checks, we include $\ebounds$ and $\enull$ as part of $\elang$
expressions which, once evaluated, result in an corresponding error state. $\eassignstack{x}{\dot a}$ is a new syntactic form that
modifies the stack variable $x$ with the result of $\dot a$. It is
essential for bounds widening. $\leq$ and $-$ are introduced
to operate on bounds and decide whether we need to halt with a bounds error
or widen a null-terminated string.

\begin{figure}[h]
{\small
  \[\begin{array}{llcl}
      \text{Atoms}       & \dot a & ::= & n \mid x \\
      
      \text{C-Expressions} & \dot c & ::= & \dot a \mid \estrlen{\dot a}  \mid \emalloc {\dot  a}  \mid \ecall{f} {\overline{\dot a}}  \\
                         & & \mid &  \mid  \dot a \circ \dot a \mid \estar{\dot a} \\
                         & & \mid & \eassign{\dot a}{\dot a} \mid \eassignstack{x}{\dot a}  \mid \eif{\dot a}{\dot e}{\dot e} \\
                         & & \mid & \ebounds \mid \enull \\
      \text{Expressions}  & \dot e & ::= & \dot c \mid \elet{x}{\dot c}{\dot e}\\
      \text{Binops} & \circ & ::= & + \mid - \mid \leq  \\
      \text{Closure} & \dot C & ::= & \hole \mid \elet{x}{\dot a}{\dot C} \\
       & & \mid & \eif{\dot a}{\dot e}{\dot C} \mid \eif{\dot a}{\dot    C}{\dot e} \\
      \text{Bounds Map} & \rho & \in & \texttt{Var} \rightharpoonup \texttt{Var} \times \texttt{Var}
  \end{array}
  \]
}
  \caption{\elang{} Syntax}
  \label{fig:syntaxerased}
{\small
  \[\begin{array}{lll}
    \dot \mu & ::= & n \mid \bot \\
    \dot c & ::= & \ldots  \mid \eret{x}{\dot \mu}{\dot e} \\
    \dot H & \in & \mathbb{Z} \rightharpoonup \mathbb{Z} \\
    \dot r & ::= & \dot e \mid \enull \mid \ebounds \\
    \dot E & ::= & \hole \mid  \elet{x}{\dot E}{\dot e}  \mid \eret{x}{i}{\dot E} \\
             & \mid & \eif{\dot E}{\dot e}{\dot e} \mid \estrlen{\dot E} \\
             & \mid & \emalloc{\dot E} \mid \ecall{f}{\overline{\dot
                      E}} \mid \dot E \circ \dot a \mid n \circ \dot E \\
             & \mid & \estar{\dot E} \mid \eassign{\dot E}{\dot a}
                      \mid \eassign{n}{\dot E} \mid
                      \eassignstack{x}{\dot E} \\
      \overline{\dot E } & ::= & \dot E \mid n,\overline{\dot E} \mid
                                 \overline{\dot E}, \dot a
                      
  \end{array} \]
}
  \caption{\elang Semantic Defs}
  \label{fig:semanticserased}
\end{figure}

$\elang$ does not include any annotations. We remove structs from
$\elang$ because we can always statically convert expressions of the form
$\eamper{n:\tau}{f}$ into $n + n_f$, where $n_f$ is the statically
determined offset of $f$ within the struct. We ellide the semantics of
$\elang$ because it is self-evident and mirrors the semantics
$\lang$. The difference is that in $\elang$, only $\ebounds$ and
$\enull$ can step into an error state. All failed dereferences and
assignments would result in a stuck state and therefore we rely on the
compiler to explicitly insert checks for checked pointers.

Fig.~\ref{fig:compilation} and Fig.~\ref{fig:compilation2}
 shows the rules for the compilation judgment for expressions,
\[\Gamma;\rho \vdash e \gg \dot C, \dot a\]
The judgment is presented differently from the one in
Sec.~\ref{sec:compilation}, which was simplified for presentation purposes. First,
we remove $\Theta$ and $m$ because these
parameters are only used for checking and have no
impact on compilation. Second, the judgment includes two
outputs, a closure $\dot C$ and an atom expression $\dot a$, instead of
a single $\elang$ expression $\dot e$. $\dot C$ can be intuitively understood as a
partially constructed program or context. Whereas $\dot E$ is used for
evaluation, $\dot C$
is used purely as a device for compilation. As an example,
when compiling $(1 : \tint) \plus (2 : \tint)$, 
we would first create a fresh variable $x$, and then produce two outputs:
\[    \dot C = \elet{x}{1 \plus 2}{\hole}\]
\[    \dot a = x\]
To obtain the compiled expression $\dot e$, we plug $\dot a$ into
$\dot C$ using the usual notation $\dot C[\dot a]$. We can also use $\dot C$ to represent runtime checks, which
usually take the form $\elet{x}{\dot c}{\hole}$, where $\dot c$
contains the check whose evaluation must not trigger $\ebounds$ or $\enull$ for the program to
continue (see Fig.~\ref{fig:compilationhelpers} for the metafunctions that create
those checks). 

This unconventional output format enables us to separate the
evaluation of the term and the computation that relies on the term's
evaluated result. Since effects and reduction (except for variables) happen only
within closures, we can precisely control the order in which effects
and evaluation happen by composing the contexts in a specific order.
Given two closures $\dot C_1$
and $\dot C_2$, we write $\dot C_1[\dot C_2]$ to denote the meta
operation of plugging $\dot C_2$ into $\dot C_1$. We also use
$\dot C_{a;b;c}$ as a shorthand for $\dot C_a[\dot C_b[\dot C_c]]$. In
the \textsc{C-Ind} rule,
we first evaluate the expressions that correspond to $e_1$ and $e_2$
through $\dot C_1$ and $\dot C_2$, and then perform a null check and
an addition through $\dot C_n$ and $\dot C_3$. Finally, we dereference
the result through $\dot C_4$ before returning the pair $\dot C_4,\dot
x_4$, propagating the flexibility to the compilation rule that recursively
calls \textsc{C-Ind}.

Fig.~\ref{fig:compilationhelpers} shows the metafunctions that create
closures representing dynamic checks. These functions first examine
whether the pointer is a checked. If the pointer is unchecked, an empty closure $\hole$
will be returned, because there is no need to perform a check. For bounds
checking, there is a special case for NT-array pointers, where the
bounds are retrived from the \textcolor{purple}{shadow} variables (found by
looking up $\rho$) on the stack  rather than
using the bounds specified in the type annotation. This is how we
achieve the same precise runtime behavior as $\lang$ in our compiled expressions.

Fig.~\ref{fig:compilationhelpers2} shows the metafunctions related to
bounds widening. $\vdash_{extend}$ takes $\rho$,
a checked NT-array pointer variable $x$, and its bounds $(b_l,b_h)$ as inputs, and returns an
extended $\rho'$ that maps $x$ to two fresh variables $x_l$, $x_h$,
together with a closure $\dot C$ that initializes $x_l$ and $x_h$ to
$b_l$ and $b_h$ respectively. This function is used in the
\textsc{C-Let} rule to extend $\rho$ before compiling the body of the
$\elettext$ binding. The updated $\rho'$ can be used for generating
precise bounds checks, and for inserting expressions that can potentially
widen the upper bounds, as seen in the $\vdash_{widenstr}$
metafunction used in the \textsc{C-Str} compilation rule.

\begin{figure*}[t]
{\small
  \begin{mathpar}
    \inferrule[]
    { x_l, x_h = \fresh \\
      \rho' = \rho[x \mapsto (x_l, x_h)]\\
    \dot C = \elet{x_l}{b_l}{\elet{x_h}{b_h}{\hole}}}
  {\cextend{\rho}{x}{\tntarrayptr{b_l}{b_h}{\tau}{c}}{\dot C, \rho'}\\}

  \inferrule[]
  {x_l,x_h = \rho(x) \\
    x_w = \fresh \\
  \dot C = \elet{x_w}{\eif{x_h}{0}{\eassignstack{x_h}{1}}}{\hole}}
{\ewidenderef{\rho}{x}{\tptr{\tntarrayb{(b_l,b_h)}{\tau}}{c}}{\dot C}}

\inferrule[]
{e \notin dom(\rho)}
{\ewidenstrlen{\rho}{e}{\dot a}{\tptr{\tntarrayb{\bvar}{\tau}}{m}}{\hole}}

\inferrule[]
{x_l,x_h = \rho(e) \\
  x_a = \fresh \\
  \dot C = \elet{x_a}{\eif{\dot a \leq x_h}{0}{\eassignstack{x_h}{\dot a}}}{\hole}}
{\ewidenstrlen{\rho}{e}{\dot a}{\tptr{\tntarrayb{\bvar}{\tau}}{\cmode}}{\dot C}}

  \end{mathpar}

}
 \caption{Metafunctions for Widening}
\label{fig:compilationhelpers2}
\end{figure*}

\begin{figure*}[h]
{\small
  \begin{mathpar}
    \inferrule[]
    {x = \fresh \\
      \dot C = \elet{x}{\eif{\dot a}{0}{\enull}}{\hole}}
    {\echecknull{\dot a}{c}{\dot C}}

    \inferrule[]
    {}
    {\echecknull{\dot a}{u}{\hole}} \\

    \inferrule[]
    {}
    {\echeckbounds{\rho}{e}{\tptr{\tallarrayb{\bvar}{\tau}}{u}}{\dot a}{\hole}} 

    \inferrule[]
    { x_l,x_h = \rho(e) \\
      x_{cl},x_{ch} = \fresh\\
      \dot C_{cl} = \elet{x_{cl}}{\eif{x_l \leq \dot a}{0}{\ebounds}}{\hole}\\
      \dot C_{ch} = \elet{x_{ch}}{\eif{\dot a \leq x_h}{0}{\ebounds}}{\hole}
    }
    {\echeckbounds{\rho}{e}{\tptr{\tallarrayb{\bvar}{\tau}}{c}}{\dot a}{\dot C_{cl;ch}}}

    \inferrule[]
    { e \notin dom(\rho) \\
      x_{l}, x_{h}, x_{cl},x_{ch} = \fresh\\
      \dot C_l = \elet{x_l}{b_l}{\hole}\\
      \dot C_h = \elet{x_h}{b_h}{\hole}  \\    
      \dot C_{cl} = \elet{x_{cl}}{\eif{x_l \leq \dot a}{0}{\ebounds}}{\hole}\\
      \dot C_{ch} = \elet{x_{ch}}{\eif{\dot a \leq x_h}{0}{\ebounds}}{\hole}
    }
    {\echeckbounds{\rho}{e}{\tntarrayptr{b_l}{b_h}{\tau}{c}}{\dot a}{\dot C_{l;h;cl;ch}}}

    \inferrule[]
    { e \notin dom(\rho) \\
      x_{l}, x_{h}, x_{cl},x_{ch} = \fresh\\
      \dot C_l = \elet{x_l}{b_l}{\hole}\\
      \dot C_h = \elet{x_h}{b_h}{\hole}  \\    
      \dot C_{cl} = \elet{x_{cl}}{\eif{x_l \leq \dot a}{0}{\ebounds}}{\hole}\\
      \dot C_{ch} = \elet{x_{ch}}{\eif{x_h \leq \dot a}{\ebounds}{0}}{\hole}
    }
    {\echeckbounds{\rho}{e}{\tarrayptr{b_l}{b_h}{\tau}{c}}{\dot a}{\dot C_{l;h;cl;ch}}}

    \inferrule[]
    {}
    {\echeckboundsw{\rho}{e}{\tptr{\tallarrayb{\bvar}{\tau}}{u}}{\dot a}{\hole}} 

    \inferrule[]
    { x_l,x_h = \rho(e) \\
      x_{cl},x_{ch} = \fresh\\
      \dot C_{cl} = \elet{x_{cl}}{\eif{x_l \leq \dot a}{0}{\ebounds}}{\hole}\\
      \dot C_{ch} = \elet{x_{ch}}{\eif{\dot a \leq x_h}{0}{\ebounds}}{\hole}
    }
    {\echeckboundsw{\rho}{e}{\tptr{\tallarrayb{\bvar}{\tau}}{c}}{\dot a}{\dot C_{cl;ch}}}

    \inferrule[]
    { e \notin dom(\rho) \\
      x_{l}, x_{h}, x_{cl},x_{ch} = \fresh\\
      \dot C_l = \elet{x_l}{b_l}{\hole}\\
      \dot C_h = \elet{x_h}{b_h}{\hole}  \\    
      \dot C_{cl} = \elet{x_{cl}}{\eif{x_l \leq \dot a}{0}{\ebounds}}{\hole}\\
      \dot C_{ch} = \elet{x_{ch}}{\eif{x_h \leq \dot a}{\ebounds}{0}}{\hole}
    }
    {\echeckboundsw{\rho}{e}{\tallarrayptr{b_l}{b_h}{\tau}{c}}{\dot a}{\dot C_{l;h;cl;ch}}}

\inferrule[]
{e \notin dom(\rho) \\
  x_l,x_l',x_h,x_h' = \fresh \\
  \dot C_1 = \elet{x_l}{b_l}{\elet{x_h}{b_h}{\hole}}\\
  \dot C_2 = \elet{x_l'}{b_l'}{\elet{x_h'}{b_h'}{\hole}}\\
  \dot C_3 = \eif{x_l' \leq x_l}{\hole}{\ebounds} \\
  \dot C_4 = \eif{x_h \leq x_h'}{\hole}{\ebounds}
}
{\echeckboundsdyn{\rho}{e}{\tptr{\tallarrayb{(b_l,b_h)}{\tau}}{m}}{\tptr{\tallarrayb{(b_l',b_h')}{\tau}}{m}}{\dot C_{1;2;3;4}}}

\inferrule[]
{x_l',x_h' =\rho(e) \\
  x_l,x_h = \fresh \\
  \dot C_1 = \elet{x_l}{b_l}{\elet{x_h}{b_h}{\hole}}\\
  \dot C_2 = \eif{x_l' \leq x_l}{\hole}{\ebounds} \\
  \dot C_3 = \eif{x_h \leq x_h'}{\hole}{\ebounds}
}
{\echeckboundsdyn{\rho}{e}{\tptr{\tallarrayb{(b_l,b_h)}{\tau}}{m}}{\tptr{\tallarrayb{(b_l',b_h')}{\tau}}{m}}{\dot C_{1;2;3}}}
  \end{mathpar}
}
\caption{Metafunctions for Dynamic Checks}
\label{fig:compilationhelpers}
\end{figure*}

\begin{figure*}[t]
{\small
  \begin{mathpar}

    \inferrule[C-Const]
      {}
      {\Gamma;\rho\vdash \evalue{n}{\tau} \gg \hole, n : \tau}

    \inferrule[C-Var]
      {x : \tau \in \Gamma}
      {\Gamma;\rho \vdash x \gg \hole, x : \tau}

    \inferrule[C-Cast]
              {\Gamma;\rho \vdash e \gg  \dot C , \dot a : \tau'}
              {\Gamma;\rho \vdash \ecast{\tau}{e} \gg \dot C, \dot a  : \tau}

      \inferrule[C-DynCast]
      {\Gamma;\rho \vdash e \gg \dot C_1, \dot a : \tptr{\tallarrayb{\bvar'}{\tau}}{m} \\
      \echeckboundsdyn{\rho}{e}{\tptr{\tallarrayb{\bvar}{\tau}}{m}}{\tptr{\tallarrayb{\bvar'}{\tau}}{m}}{\dot C_{b}}}
      {\Gamma;\rho \vdash \edyncast{\tptr{\tallarrayb{\bvar}{\tau}}{m}}{e}
            \gg \dot C_{1;b}, \dot a : \tptr{\tallarrayb{\bvar}{\tau}}{m}}

      \inferrule[C-Str]
      {\Gamma; \rho \vdash e \gg \dot C_1, \dot a_1 :  \tptr{\tntarrayb{\bvar}{\tau_a}}{m} \\
        \echecknull{\dot a_1}{m}{\dot C_{n}} \\
        \echeckbounds{\rho}{\dot a_1}{\tptr{\tntarrayb{\bvar}{\tau_a}}{m}}{0}{\dot C_{b}} \\
        x_2 = \fresh \\
          \dot C_2 = \elet{x_2}{\estrlen{\dot a_1}}{\hole} \\
      \ewidenstrlen{\rho}{e}{\dot a_1}{\tptr{\tntarrayb{\bvar}{\tau_a}}{m}}{\dot C_{w}} } 
                {\Gamma;\rho \vdash \estrlen{e} \gg \dot C_{1;n;b;2;w}, x_2 : \tint}

      \inferrule[C-LetStr]
      {\Gamma(y) = \tntarrayptr{b_l}{b_h}{\tau_a}{\cmode} \\ x \not\in \fv(\tau) \\
        \Gamma; \rho \vdash \estrlen{y} \gg \dot C_1, \dot a_1 : \tint\\
        \dot C_2 = \elet{x}{\dot a_1}{\hole} \\
        \Gamma[x\mapsto \tint,y\mapsto [\tntarrayptr{b_l}{x}{\tau_a}{\cmode}]];\rho \vdash e_3 \gg \dot C_3, \dot a_3 : \tau}
                {\Gamma;\rho \vdash \elet{x}{\estrlen{y}}{e} \gg \dot C_{1;2;3},\dot a_3 : \tau}

      \inferrule[C-If ]
      {\Gamma; \rho \vdash e \gg \dot C_1, \dot a_1 : \tau \\
        \Gamma; \rho \vdash e_1 \gg \dot C_2, \dot a_2 : \tau_2 \\
        \Gamma; \rho \vdash e_3 \gg \dot C_3, \dot a_3 : \tau_3 \\
        x_4 = \fresh \\
        \dot C_4 = \elet{x_4}{\eif{\dot a_1}{\dot C_2[\dot a_2]}{\dot C_3[\dot a_3]}}{\hole}
      }
        {\Gamma; \rho \vdash \eif{e_1}{e_2}{e_3} \gg \dot C_{1;4}, x_4 : \tau_2 \sqcup \tau_3}

        \inferrule[C-IfNT]
        {\Gamma; \rho \vdash x : \tptr{\tntarrayb{(b_l,b_h)}{\tau}}{c} \\
          b_h = 0 \Rightarrow \Gamma' = \Gamma[x \mapsto \tptr{\tntarrayb{(b_l,1)}{\tau}}{c}] \\
          b_h \neq 0 \Rightarrow \Gamma' = \Gamma \\
          \Gamma; \rho \vdash \estar{x} \gg \dot C_1, \dot a_1 : \tau_1\\
          \Gamma'; \rho \vdash e_2 \gg \dot C_2, \dot a_2 : \tau_2\\
          \Gamma;\rho \vdash e_3 \gg \dot C_3, \dot a_3 : \tau_3 \\
        \ewidenderef{\rho}{x}{\tptr{\tntarrayb{(b_l,b_h)}{\tau}}{c}}{\dot C_{w}} \\
        x_4 = \fresh \\
        \dot C_4 = \elet{x_4}{\eif{\dot a_1}{\dot C_{2;w}[\dot a_2]}{\dot C_3[\dot a_3]}}{\hole}}
        {\Gamma;\rho \vdash \eif{\estar{x}}{e_1}{e_2} \gg \dot C_{1;4}, x_4 : \tau_1 \sqcup \tau_2}

   \inferrule[C-Let]
    { (x\in \fv(\tau') \Rightarrow e_1 \in Bound) \\
        \Gamma;\rho \vdash e_1 \gg \dot C_1, \dot a_1 : \tau_1 \\ 
        \cextend{\rho}{x}{\tau_1}{\dot C_2,\rho'}\\
        \dot C_3 = \elet{x}{\dot a_1}{\hole}\\
        \Gamma[x\mapsto \tau];\rho' \vdash e_4 \gg \dot C_4, \dot a_4  : \tau_4}
    {\Gamma;\rho' \vdash \elet{x}{e_1}{e_4} \gg \dot C_{1;2;3;4},\dot a_4 : \tau_4[\tau_1 = \tint \Rightarrow x \mapsto e_1]}

  \inferrule[C-Ret]
    {\Gamma(x)\neq \bot \\
      \Gamma;\rho \vdash e \gg \dot C_1, \dot a_1 : \tau \\
      x_2 = \fresh \\
      \mu \gg \dot \mu \\
    \dot C_2 = \elet{x_2}{\eret{x}{\dot \mu}{\dot C_1[\dot a_1]}}{\hole}}
  {\Gamma;\rho \vdash \eret{x}{\mu}{e} \gg \dot C_2, x_2 : \tau}

    \inferrule[C-Fun]
    {\Xi(f) = \tau\;(\overline{x}:\overline{\tau})\;e \\
      (\forall e_i\in\overline{e}\;\;\tau_i\in\overline{\tau}\;.\; \Gamma; \rho \vdash {e_i} \gg \dot C_i, \dot a_i  : \\ \tau_i' \wedge
      \tau_i' \sqsubseteq 
      \tau_i[\overline{e} / \overline{x}]
      ) \\
      x_f = \fresh \\
      \dot C_f = \elet{x_f}{f(\overline{a})}{\hole}}
    {\Gamma; \rho \vdash f(\overline{e}) \gg \overline{\dot C}[\dot C_f] , x_f  : \tau[\overline{e} / \overline{x}]}

  \inferrule[C-Def]
  {\Gamma;\rho \vdash e_1 \gg \dot C_1, \dot a_1 : \tptr{\tau}{m} \\
    \echecknull{\dot a_1}{m}{\dot C_{n}}\\
    x_2 = \fresh \\
    \dot C_2 = \elet{x_2}{\estar{\dot a_1}}{\hole}
   }
   {\Gamma;\rho \vdash \estar{e_1} \gg \dot C_{1;n;2}, x_2 : \tau}

   \inferrule[C-DefArr]
   {\Gamma;\rho \vdash e_1 \gg \dot C_1, \dot a_1 : \tallarrayptr{b_l}{b_h}{\tau}{m} \\
    \echecknull{\dot a_1}{m}{\dot C_{n}}\\
    \echeckbounds{\rho}{e_1}{\tallarrayptr{b_l}{b_h}{\tau}{m}}{0}{\dot C_{b}}\\
    x_2 = \fresh \\
    \dot C_2 = \elet{x_2}{\estar{\dot a_1}}{\hole}}
   {\Gamma;\rho \vdash \estar{e_1} \gg \dot C_{1;n;b;2},x_2 : \tau}

    \inferrule[C-Mac]
    {\dot C_{1}, \dot a_1 = \esizeof{\omega} \\
      x_2 = \fresh \\
    \dot C_2 = \elet{x_2}{\emalloc{\dot a_1}}{\hole}}
     {\Gamma; \rho \vdash \emalloc{\omega} \gg \dot C_{1;2}, x_2 : \tptr{\omega}{\cmode}}

  \end{mathpar}
}
\caption{Compilation}
\label{fig:compilation}
\end{figure*}

\begin{figure*}[t]
{\small
  \begin{mathpar}
    \inferrule[C-Add]
              {\Gamma; \rho \vdash e_1 \gg \dot C_1, \dot a_1 : \tint \\
                \Gamma; \rho \vdash e_2 \gg \dot C_2, \dot a_2 : \tint\\
                x_3 = \fresh \\
              \dot C_3 = \elet{x_3}{\dot a_1 \plus \dot a_2}{\hole}}
              {\Gamma; \rho \vdash \dot C_3, x_3 : \tint }

   \inferrule[C-Ind] 
              {\Gamma; \rho \vdash e_1 \gg \dot C_1, \dot a_1 : \tptr{\tallarrayb{\bvar}{\tau}}{m} \\
                \Gamma; \rho \vdash e_2 \gg \dot C_2, \dot a_2 : \tint \\
                \echecknull{\dot a_1}{m}{\dot C_{n}}\\
                \echeckbounds{\rho}{e_1}{\tptr{\tallarrayb{\bvar}{\tau}}{m}}{\dot a_2}{\dot C_{b}} \\
                x_3, x_4 = \fresh\\
                \dot C_3 = \elet{x_3}{\dot a_1 \plus \dot a_2}{\hole}\\
                \dot C_4 = \elet{x_4}{\estar{x_3}}{\hole}
                }
                {\Gamma; \rho \vdash \estar{(\ebinop{e_1}{e_2})} \gg \dot C_{1;2;n;3;b;4} , x_4: \tau}

  \inferrule[C-Assign]
  {\Gamma; \rho \vdash e_1 \gg \dot C_1, \dot a_1 : \tptr{\tau}{m'} \\
    \echecknull{\dot a_1}{m}{\dot C_{n}}\\
    \Gamma; \rho \vdash e_2 \gg \dot C_2, \dot a_2 : \tau' \\
    \tau'\sqsubseteq \tau \\
    x_3 = \fresh \\
    \dot C_3 = \elet{x_3}{\eassign{\dot a_1}{\dot a_2}}{\hole} \\
  }
  {\Gamma; \rho \vdash \eassign{e_1}{e_2} \gg \dot C_{1;2;n;3}, x_3 : \tau}

  \inferrule[C-AssignArr]
  {\Gamma; \rho \vdash e_1 \gg \dot C_1, \dot a_1 : \tptr{\tallarrayb{\bvar}{\tau}}{m'}\\
    \echecknull{\dot a_1}{m}{\dot C_{n}}\\
    \echeckboundsw{\rho}{e_1}{\tallarrayptr{b_l}{b_h}{\tau}{m}}{0}{\dot C_{b}}\\
    \Gamma; \rho \vdash e_2 \gg \dot C_2, \dot a_2 : \tau' \\
    x_3 = \fresh \\
    \dot C_3 = \elet{x_3}{\eassign{\dot a_1}{\dot a_2}}{\hole} \\
    \tau'\sqsubseteq \tau}
  {\Gamma; \rho \vdash \eassign{e_1}{e_2} \gg \dot C_{1;2;n;b;3},x_3 : \tau}

    \inferrule[C-IndAssign]
    {\Gamma; \rho \vdash e_1 \gg \dot C_1, \dot a_1 : \tptr{\tallarrayb{\bvar}{\tau}}{m} \\
      \Gamma; \rho \vdash e_2 \gg \dot C_2, \dot a_2 : \tint \\
      \echecknull{\dot a_1}{m}{\dot C_{n}}\\
      \echeckboundsw{\rho}{e_1}{\tptr{\tallarrayb{\bvar}{\tau}}{m}}{\dot a_2}{\dot C_{b}} \\
      \Gamma; \rho \vdash e_3 \gg \dot C_3, \dot a_3 : \tau' \\
      x_4, x_5 = \fresh \\
      \dot C_4 = \elet{x_4}{\dot a_1 \plus \dot a_2}{\hole}\\
      \dot C_5 = \elet{x_5}{\eassign{x_4}{x_3}{\hole}}
      \tau'\sqsubseteq \tau}
    {\Gamma; \rho \vdash \eassign{(e_1 \plus e_2)}{e_3} \gg \dot C_{1;2;n;3;4;b;5} : \tau}

    \inferrule [C-Struct]
    {\Gamma; \rho \vdash e_1 \gg \dot C_1, \dot a_1  :  {\tptr{\tstruct{T}}{m}} \\
      D(T) = \tau_0~f_0\ldots; \tau_j~f; ...  \\
      \echecknull{\dot a_1}{m}{\dot C_{n}}\\
      x_2 = \fresh\\
    \dot C_2 = \elet{x_2}{\dot a_1 \plus \dot j}{\hole}}
    {\Gamma; \rho \vdash \eamper{e_1}{f} \gg \dot C_2, x_2  : \tptr{\tau_f}{m}}

    \inferrule[C-Unchecked]
              {\Gamma;\rho \vdash e \gg \dot C, \dot a : \tau}
              {\Gamma;\rho \vdash \eunchecked{e} \gg \dot C, \dot a : \tau}

  \end{mathpar}
}
\caption{Compilation (Continued)}
\label{fig:compilation2}
\end{figure*}

%% file: main.bbl
\begin{thebibliography}{28}
\providecommand{\natexlab}[1]{#1}
\providecommand{\url}[1]{\texttt{#1}}
\expandafter\ifx\csname urlstyle\endcsname\relax
  \providecommand{\doi}[1]{doi: #1}\else
  \providecommand{\doi}{doi: \begingroup \urlstyle{rm}\Url}\fi

\bibitem[Blazy and Leroy(2009)]{Blazy2009}
Sandrine Blazy and Xavier Leroy.
\newblock {Mechanized Semantics for the Clight Subset of the C Language}.
\newblock \emph{Journal of Automated Reasoning}, 43\penalty0 (3):\penalty0
  263--288, 2009.
\newblock ISSN 1573-0670.
\newblock \doi{10.1007/s10817-009-9148-3}.
\newblock URL \url{http://dx.doi.org/10.1007/s10817-009-9148-3}.

\bibitem[Condit et~al.(2007)Condit, Harren, Anderson, Gay, and
  Necula]{Condit2007}
Jeremy Condit, Matthew Harren, Zachary Anderson, David Gay, and George~C.
  Necula.
\newblock {Dependent Types for Low-Level Programming}.
\newblock In \emph{Proceedings of European Symposium on Programming (ESOP
  '07)}, 2007.

\bibitem[Duan et~al.(2020)Duan, Yang, Zhou, and Criswell]{duanrefactoring}
Junhan Duan, Yudi Yang, Jie Zhou, and John Criswell.
\newblock {Refactoring the FreeBSD Kernel with Checked C}.
\newblock In \emph{IEEE Cybersecurity Development Conference (SecDev)},
  September 2020.

\bibitem[Elliott et~al.(2018)Elliott, Ruef, Hicks, and Tarditi]{Elliott2018}
Archibald~Samuel Elliott, Andrew Ruef, Michael Hicks, and David Tarditi.
\newblock {Checked C: Making C Safe by Extension}.
\newblock In \emph{2018 IEEE Cybersecurity Development (SecDev)}, pages 53--60,
  2018.
\newblock \doi{10.1109/SecDev.2018.00015}.

\bibitem[Ellison and Rosu(2012)]{ellison-rosu-2012-popl}
Chucky Ellison and Grigore Rosu.
\newblock {An Executable Formal Semantics of C with Applications}.
\newblock In \emph{Proceedings of the 39th Annual ACM SIGPLAN-SIGACT Symposium
  on Principles of Programming Languages}, POPL '12, pages 533--544, New York,
  NY, USA, 2012. ACM.
\newblock ISBN 978-1-4503-1083-3.
\newblock \doi{10.1145/2103656.2103719}.
\newblock URL \url{http://doi.acm.org/10.1145/2103656.2103719}.

\bibitem[Felleisen et~al.(2009)Felleisen, Findler, and Flatt]{pltredex}
Matthias Felleisen, Robert~Bruce Findler, and Matthew Flatt.
\newblock \emph{{Semantics Engineering with PLT Redex}}.
\newblock The MIT Press, 1st edition, 2009.
\newblock ISBN 0262062755.

\bibitem[Grossman et~al.(2002)Grossman, Morrisett, Jim, Hicks, Wang, and
  Cheney]{GrossmanMJHWC02}
Dan Grossman, Greg Morrisett, Trevor Jim, Michael Hicks, Yanling Wang, and
  James Cheney.
\newblock {Region-based Memory Management in {C}yclone}.
\newblock In \emph{PLDI}, 2002.

\bibitem[Guha et~al.(2010)Guha, Saftoiu, and Krishnamurthi]{lambdajs}
Arjun Guha, Claudiu Saftoiu, and Shriram Krishnamurthi.
\newblock {The Essence of Javascript}.
\newblock In \emph{Proceedings of the 24th European Conference on
  Object-Oriented Programming}, ECOOP'10, page 126–150, Berlin, Heidelberg,
  2010. Springer-Verlag.
\newblock ISBN 3642141064.

\bibitem[Hathhorn et~al.(2015)Hathhorn, Ellison, and
  Ro\c{s}u]{10.1145/2813885.2737979}
Chris Hathhorn, Chucky Ellison, and Grigore Ro\c{s}u.
\newblock {Defining the Undefinedness of C}.
\newblock \emph{SIGPLAN Not.}, 50\penalty0 (6):\penalty0 336–345, June 2015.
\newblock ISSN 0362-1340.
\newblock \doi{10.1145/2813885.2737979}.
\newblock URL \url{https://doi.org/10.1145/2813885.2737979}.

\bibitem[Jim et~al.(2002)Jim, Morrisett, Grossman, Hicks, Cheney, , and
  Wang]{Jim2002}
Trevor Jim, Greg Morrisett, Dan Grossman, Michael Hicks, James Cheney, , and
  Yanling Wang.
\newblock {Cyclone: A Safe Dialect of {C}}.
\newblock In \emph{USENIX Annual Technical Conference}, pages 275--288,
  Monterey, CA, 2002. {USENIX}.

\bibitem[Kang et~al.(2015)Kang, Hur, Mansky, Garbuzov, Zdancewic, and
  Vafeiadis]{Kang:2015:FCM:2813885.2738005}
Jeehoon Kang, Chung-Kil Hur, William Mansky, Dmitri Garbuzov, Steve Zdancewic,
  and Viktor Vafeiadis.
\newblock {A Formal C Memory Model Supporting Integer-pointer Casts}.
\newblock \emph{SIGPLAN Not.}, 50\penalty0 (6):\penalty0 326--335, June 2015.
\newblock ISSN 0362-1340.
\newblock \doi{10.1145/2813885.2738005}.
\newblock URL \url{http://doi.acm.org/10.1145/2813885.2738005}.

\bibitem[Lampropoulos and Pierce(2018)]{Pierce:SF4}
Leonidas Lampropoulos and Benjamin~C. Pierce.
\newblock \emph{{QuickChick}: Property-Based Testing in Coq}.
\newblock Software Foundations series, volume 4. Electronic textbook, August
  2018.
\newblock Version 1.0. \url{http://www.cis.upenn.edu/~bcpierce/sf}.

\bibitem[Leroy(2009)]{compcert}
Xavier Leroy.
\newblock Formal verification of a realistic compiler.
\newblock \emph{Communications of the ACM}, 52\penalty0 (7):\penalty0 107--115,
  July 2009.
\newblock ISSN 0001-0782.
\newblock \doi{10/c9sb7q}.
\newblock URL \url{http://doi.acm.org/10.1145/1538788.1538814}.

\bibitem[Leroy et~al.(2012)Leroy, Appel, Blazy, and
  Stewart]{leroy:hal-00703441}
Xavier Leroy, Andrew~W. Appel, Sandrine Blazy, and Gordon Stewart.
\newblock {The CompCert Memory Model, Version 2}.
\newblock Research Report RR-7987, {INRIA}, June 2012.
\newblock URL \url{https://hal.inria.fr/hal-00703441}.

\bibitem[Memarian et~al.(2016)Memarian, Matthiesen, Lingard, Nienhuis,
  Chisnall, Watson, and Sewell]{10.1145/2980983.2908081}
Kayvan Memarian, Justus Matthiesen, James Lingard, Kyndylan Nienhuis, David
  Chisnall, Robert N.~M. Watson, and Peter Sewell.
\newblock {Into the Depths of C: Elaborating the de Facto Standards}.
\newblock \emph{SIGPLAN Not.}, 51\penalty0 (6):\penalty0 1–15, June 2016.
\newblock ISSN 0362-1340.
\newblock \doi{10.1145/2980983.2908081}.
\newblock URL \url{https://doi.org/10.1145/2980983.2908081}.

\bibitem[Memarian et~al.(2019)Memarian, Gomes, Davis, Kell, Richardson, Watson,
  and Sewell]{Memarian:2019:ECS:3302515.3290380}
Kayvan Memarian, Victor B.~F. Gomes, Brooks Davis, Stephen Kell, Alexander
  Richardson, Robert N.~M. Watson, and Peter Sewell.
\newblock {Exploring C Semantics and Pointer Provenance}.
\newblock \emph{Proc. ACM Program. Lang.}, 3\penalty0 (POPL):\penalty0
  67:1--67:32, January 2019.
\newblock ISSN 2475-1421.
\newblock \doi{10.1145/3290380}.
\newblock URL \url{http://doi.acm.org/10.1145/3290380}.

\bibitem[Merigoux et~al.(2021)Merigoux, Chataing, and
  Protzenko]{merigoux2021catala}
Denis Merigoux, Nicolas Chataing, and Jonathan Protzenko.
\newblock {Catala: A Programming Language for the Law}.
\newblock \emph{arXiv preprint arXiv:2103.03198}, 2021.

\bibitem[Nagarakatte et~al.(2009)Nagarakatte, Zhao, Martin, and
  Zdancewic]{softbound}
Santosh Nagarakatte, Jianzhou Zhao, Milo~M.K. Martin, and Steve Zdancewic.
\newblock {SoftBound: Highly Compatible and Complete Spatial Memory Safety for
  C}.
\newblock In \emph{Proceedings of the 30th ACM SIGPLAN Conference on
  Programming Language Design and Implementation}, PLDI '09, page 245–258,
  New York, NY, USA, 2009. Association for Computing Machinery.
\newblock ISBN 9781605583921.
\newblock \doi{10.1145/1542476.1542504}.
\newblock URL \url{https://doi.org/10.1145/1542476.1542504}.

\bibitem[{Necula} et~al.(2005){Necula}, Condit, Harren, McPeak, and
  Weimer]{Necula2005}
George~C. {Necula}, Jeremy Condit, Matthew Harren, Scott McPeak, and Westley
  Weimer.
\newblock {{CCured}: Type-Safe Retrofitting of Legacy Software}.
\newblock \emph{{ACM} Transactions on Programming Languages and Systems
  ({TOPLAS})}, 27\penalty0 (3), 2005.

\bibitem[Pa{\l}ka et~al.(2011)Pa{\l}ka, Claessen, Russo, and
  Hughes]{PalkaAST11}
Micha{\l}~H. Pa{\l}ka, Koen Claessen, Alejandro Russo, and John Hughes.
\newblock {Testing an Optimising Compiler by Generating Random Lambda Terms}.
\newblock In \emph{Proceedings of the 6th International Workshop on Automation
  of Software Test}, AST '11, pages 91--97, New York, NY, USA, 2011. ACM.
\newblock ISBN 978-1-4503-0592-1.
\newblock \doi{10.1145/1982595.1982615}.
\newblock URL \url{http://doi.acm.org/10.1145/1982595.1982615}.

\bibitem[Peña(2017)]{LiquidHaskell}
Ricardo Peña.
\newblock {An Introduction to Liquid Haskell}.
\newblock \emph{Electronic Proceedings in Theoretical Computer Science},
  237:\penalty0 68–80, Jan 2017.
\newblock ISSN 2075-2180.
\newblock \doi{10.4204/eptcs.237.5}.
\newblock URL \url{http://dx.doi.org/10.4204/EPTCS.237.5}.

\bibitem[Ruef et~al.(2019)Ruef, Lampropoulos, Sweet, Tarditi, and
  Hicks]{ruef18checkedc-incr}
Andrew Ruef, Leonidas Lampropoulos, Ian Sweet, David Tarditi, and Michael
  Hicks.
\newblock {Achieving Safety Incrementally with Checked C}.
\newblock In Flemming Nielson and David Sands, editors, \emph{Principles of
  Security and Trust}, pages 76--98, Cham, 2019. Springer International
  Publishing.
\newblock ISBN 978-3-030-17138-4.

\bibitem[Serebryany et~al.(2012)Serebryany, Bruening, Potapenko, and
  Vyukov]{Serebryany2012}
Konstantin Serebryany, Derek Bruening, Alexander Potapenko, and Dmitry Vyukov.
\newblock {{AddressSanitizer}: A Fast Address Sanity Checker}.
\newblock In \emph{Proceedings of the 2012 USENIX Conference on Annual
  Technical Conference}, 2012.

\bibitem[Tarditi(2021)]{checkedc}
David Tarditi.
\newblock {Extending C with Bounds Safety and Improved Type Safety}, 2021.
\newblock URL \url{https://github.com/secure-sw-dev/checkedc/}.

\bibitem[Vazou et~al.(2014)Vazou, Seidel, Jhala, Vytiniotis, and
  Peyton-Jones]{Refinementlh}
Niki Vazou, Eric~L. Seidel, Ranjit Jhala, Dimitrios Vytiniotis, and Simon
  Peyton-Jones.
\newblock {Refinement Types for Haskell}.
\newblock \emph{SIGPLAN Not.}, 49\penalty0 (9):\penalty0 269–282, August
  2014.
\newblock ISSN 0362-1340.
\newblock \doi{10.1145/2692915.2628161}.
\newblock URL \url{https://doi.org/10.1145/2692915.2628161}.

\bibitem[Zeng et~al.(2013)Zeng, Tan, and
  Erlingsson]{Zeng:2013:SRF:2534766.2534798}
Bin Zeng, Gang Tan, and \'{U}lfar Erlingsson.
\newblock {Strato: A Retargetable Framework for Low-level Inlined-reference
  Monitors}.
\newblock In \emph{Proceedings of the 22Nd USENIX Conference on Security},
  2013.

\bibitem[Zhao et~al.(2012)Zhao, Nagarakatte, Martin, and
  Zdancewic]{Zhao:2012:FLI:2103621.2103709}
Jianzhou Zhao, Santosh Nagarakatte, Milo~M.K. Martin, and Steve Zdancewic.
\newblock {Formalizing the LLVM Intermediate Representation for Verified
  Program Transformations}.
\newblock \emph{SIGPLAN Not.}, 47\penalty0 (1):\penalty0 427--440, January
  2012.
\newblock ISSN 0362-1340.
\newblock \doi{10.1145/2103621.2103709}.
\newblock URL \url{http://doi.acm.org/10.1145/2103621.2103709}.

\bibitem[Zhou et~al.(2006)Zhou, Condit, Anderson, Bagrak, Ennals, Harren,
  Necula, and Brewer]{Feng2006}
Feng Zhou, Jeremy Condit, Zachary Anderson, Ilya Bagrak, Rob Ennals, Matthew
  Harren, George Necula, and Eric Brewer.
\newblock {SafeDrive}: Safe and recoverable extensions using language-based
  techniques.
\newblock In \emph{7th Symposium on Operating System Design and Implementation
  (OSDI'06)}, Seattle, Washington, 2006. USENIX Association.

\end{thebibliography}
